# Dynamic Tracking Error and the Total Portfolio Approach

Ashwin Alankar, Allan Maymin, Philip Maymin, Myron Scholes, and Sujiang Zhang

**Corresponding Author:** Philip Maymin, philip.maymin@janushenderson.com,

Janus Henderson Investors, 151 Detroit Street, Denver, CO 80206.

**Disclosure Statement (Conflicts of Interest):** All authors are employees of Janus Henderson Investors. Janus Henderson Investors offers investment products and services, including dynamic asset allocation strategies, that implement frameworks in the class discussed in this paper. The authors have a financial interest in the commercial outcomes of Janus Henderson Investors. The opinions and analyses presented are the authors' own and do not necessarily reflect those of Janus Henderson Investors. No forecasts can be guaranteed.

# Dynamic Tracking Error and the Total Portfolio Approach

**Abstract:** Strategic Asset Allocation and the Total Portfolio Approach differ in one thing: the tracking error the board grants the chief investment officer. The board's first decision should be the drawdown it can tolerate; the benchmark and tracking error budget follow. The value comes from spending that budget dynamically, adding active risk when the reward is high and shedding it as the fund nears its limit. Managed this way, a fund would have compounded about two percentage points a year faster at the same drawdown, with no security selection. What matters is dynamic versus static tracking error, not the label.

## 1. Introduction

The institutional question is not whether to adopt TPA or SAA. It is how much discretion to grant the chief investment officer, over what, and on what time scale. Under Strategic Asset Allocation the board fixes asset-class weights and assigns each silo a narrow tracking error budget, and the total-fund tracking error is whatever falls out. Under the Total Portfolio Approach the board names one benchmark, sets one total tracking error budget, and asks the CIO to use it. The board owns the benchmark and the budget; the CIO owns everything else. Almost every reported difference between TPA and SAA portfolios follows from that one reassignment of authority (Thinking Ahead Institute 2019; CAIA Association 2024; CalPERS 2025).

Read as governance technology, the two are steps on a single path, and the path is drawdown control. SAA was the first step: fix the weights, cap each sleeve, and the fund's worst year is whatever the weights imply. TPA is the second: name one benchmark, grant one budget, and let the CIO vary risk as conditions change. The reason to widen the budget is to manage drawdown better than fixed weights can, adding risk when the reward is high and holding back when it is not. The step this paper describes is to make that goal explicit and budget the drawdown itself, treating tracking error, even a dynamic one, as a proxy for the quantity the board actually cares about. Seen this way TPA is a step beyond SAA rather than its rival, and dynamic risk management is the capability that step requires.

The board-CIO relationship carries the weight of the transfer. We treat the first mission of any board as three connected questions: what average risk level, and why is it right for the beneficiaries; how often that level changes, and by what mechanism; who decides, with what evidence. Under SAA these split cleanly across asset classes and review on a long cadence. Under TPA they sit at the total-fund level and must be answered in near real time, against the tension between governance that is by design slow and opportunities that are by nature fast. The board cannot delegate the mission without losing the substance of governance, nor retain it in full without losing the substance of TPA. The framework lives in where that line is drawn.

The budget is not set in tracking error units in the first instance. The board begins with the maximum drawdown the fund can sustain before it would be forced to act, and that limit defines the benchmark and the tracking error the CIO may run against it. A board that allows a one-year drawdown near 30%, against equities that fell roughly 50% in 2008, is implicitly sanctioning something close to a 60/40 benchmark, and the tracking error budget is what remains of the drawdown allowance once the benchmark takes its share. The fund is then a base portfolio of broad equities, duration, and cash plus a tracking error portfolio, and both draw on the same drawdown capacity. In a crisis their losses correlate, so the prudent working assumption is that their drawdowns add. We formalize this in Section 4.5; it is the lens through which the rest of the paper reads.

What makes the boundary worth drawing is the economic content on the CIO's side, and here two things are often conflated. The usual case for TPA is *concentration alpha*: overweighting attractive opportunities and giving up diversification for expected reward. Concentration alpha is real but runs into estimation error, liquidity limits, peer pressure, and the career risk of standing far from the crowd through a drawdown. The more durable component is *dynamic alpha*: varying the amount of active risk with the opportunity set, which earns a compound advantage from the same arithmetic that makes a sequence of moderate bets beat one large bet, and which needs no security-selection skill. Dynamic alpha is also often confused with cross-sectional rotation, swapping cheap exposures for rich ones inside a sleeve; only time-series scaling of total active risk compounds, and that is the claim of this paper. What such a policy maximizes is compound return, not a single-period ratio. The Sharpe ratio is the wrong yardstick for it: an active bet collinear with the benchmark, as ours is, cannot raise the Sharpe ratio because it only slides the fund along the line it already sits on. The scorecard that registers a sizing decision is compound return per unit of the drawdown the board has approved.

The harder problem is that TPA mandates drift back toward SAA with no one deciding to make the change. A CIO granted a wide budget who takes a meaningful drawdown finds it renegotiated, the next CIO inherits a narrower de facto budget, and after two or three cycles the fund runs a realized tracking error indistinguishable from a well-run SAA shop while the documents still say TPA. The three pillars of alignment, integration, and adaptability that organize recent practice (Mercer 2026) are necessary but not sufficient: none says how the tracking error policy should move through a cycle. Mercer defines adaptability as cross-sectional competition for capital and disclaims market timing, leaving untouched the time-series scaling our policy occupies, and its integration pillar carries a group-think cost, since one team that votes on everything loses independence. The convergence we document later is consistent with this drift.

We make four contributions. First, the board's primary control is the drawdown budget and the tracking error budget is its residual; a tracking error portfolio that diversifies the base in a crisis or sheds risk as the fund nears its limit frees drawdown budget for the base, raising compound return at the drawdown the board already approved (Section 4.5). Second, Sharpe ratios are statistically flat across the full range of budgets, because a bet collinear with the benchmark cannot manufacture risk-adjusted return; the value of dynamic tracking error is compound return per unit of drawdown, which the Sharpe ratio cannot see. Third, we identify *omega*, the implicit cost of the institutional tracking error limit, as the object linking governance, delegated management, and dynamic portfolio choice; it is countercyclical and large, and it is a return premium for bearing crisis risk rather than a Sharpe premium. Fourth, the part of a TPA mandate worth keeping is dynamic risk management, not concentration, and it requires no security selection; our reconciliation with the volatility-managed portfolios of Moreira and Muir (2017, 2019) shows their signal cuts on the rising-volatility leg and ours adds on the descending leg, the two being sequential complements.

## 2. Related Literature

The paper sits at the intersection of three literatures: portfolio delegation and tracking error constraints, dynamic portfolio choice, and institutional governance.

### 2.1 Tracking Error, Benchmarks, and Delegated Management

The mean-variance framework underlying both SAA and TPA is Markowitz (1952). The case for SAA rests on Brinson, Hood, and Beebower (1986) and Brinson, Singer, and Beebower (1991), who found allocation policy explains more than 90% of the variation in pension fund returns, with Ibbotson and Kaplan (2000) separating the time-series and cross-sectional shares. That work fixed attention on the level of allocation; our object is its variation over time. Roll (1992) and Jorion (2003) show a tight tracking error constraint induces benchmark-hugging and a tilt toward higher-beta securities, and Leibowitz and Henriksson (1988) study the parallel surplus problem. Treynor and Black (1973), Grinold (1989), and Grinold and Kahn (2000) derived the optimal active weight and the Fundamental Law of Active Management. Our model adds the time dimension: when the opportunity set varies across regimes, allocating the risk budget across time becomes a first-order source of value, distinct from the cross-sectional breadth the Fundamental Law emphasizes.

### 2.2 Dynamic Portfolio Choice

Merton (1969, 1971) established that when opportunities change over time the optimal portfolio differs from the single-period mean-variance solution, and Campbell and Viceira (2002) brought this to practical allocation. Our H3 is the active-management analogue: a CIO who varies tracking error with the regime does for active risk what Merton's investor does for total risk. Ang (2014) extended factor investing to total-portfolio management but treats the cross-sectional structure as static. The market-timing literature (Henriksson and Merton 1981) evaluates directional forecasts of the market; our object is the magnitude of active risk through time, not the sign of a market call.

### 2.3 Liquidity, Constraints, and Crisis Dynamics

Brunnermeier and Pedersen (2009) showed margin constraints create destabilizing liquidity spirals, and He and Krishnamurthy (2013) formalized the channel through intermediary asset pricing; our omega is analogous, binding when the cost of binding is greatest. Scholes (2000) read LTCM as forced liquidation. Alankar, Blaustein, and Scholes (2013) extended the idea to delegated management, proving that tracking error and liquidity constraints jointly force mean-variance-inefficient portfolios and identifying the joint shadow price, which they term omega, as a persistent hedging demand. Our use of omega draws on their formalization.

### 2.4 Pension Governance and the TPA Movement

Ambachtsheer (2007, 2016) documented that funds with more effective governance outperform peers. Ang, Goetzmann, and Schaefer (2009) showed Norway's fund earned its active value-add principally through systematic factor exposures, anticipating our argument that the tracking error constraint, not the organizational structure, is the binding governance input. The Thinking Ahead Institute (2019) and CAIA Association (2024) document TPA practice across leading investors, with the caveat that adopters are self-selected: the sovereign mandates and long horizons of CPP Investments, New Zealand Superannuation, and the Australian Future Fund (Future Fund 2024) differ from typical U.S. public plans. Our contribution is to reframe the TPA-versus-SAA question as a difference in the realized tracking error policy function, its level, its variation, and its cyclicality, and to establish a convergence result: under realistic governance pressures the two become empirically indistinguishable.

## 3. The Governance of Tracking Error

### 3.1 Two Modes of Governance

SAA and TPA share one objective: maximize the compound growth of the fund subject to the drawdown the board can tolerate. The board's primary instrument is the drawdown limit, and the benchmark and the tracking error budget follow from it. The two differ only in how the governance mechanism constrains the CIO, that is, in the width and rigidity of the budget the drawdown limit permits.

Under SAA the board governs through constraints. Asset-class allocations are fixed and the CIO's discretion operates within them; each sleeve carries a tight tracking error budget of perhaps 50 to 150 basis points (Jorion 2003), and total-fund tracking error emerges as a byproduct rather than a managed quantity. Under TPA the board governs through trust. The CIO receives discretion over the whole portfolio with a total-fund budget of perhaps 3% to 5% (CAIA Association 2024), can move capital across asset classes and vary risk over time, and the board must understand the process well enough to hold confidence through inevitable underperformance. Trust and constraints are substitutes: as trust erodes, after a drawdown or a change in board composition, the natural response is to reimpose constraints, which makes TPA inherently less stable than SAA.

### 3.2 The Asymmetry of Trust

Trust between a board and a CIO accumulates slowly and collapses quickly. Three years of outperformance earn incremental confidence; one quarter down 500 basis points can lose it overnight, because good outcomes do not distinguish skill from luck while a large loss reveals concretely that the CIO was willing to take risk of that size. A 5% budget is available in principle, but using it fully exposes the CIO to career risk the budget does not compensate, so the rational response is self-censorship: operate well inside the constraint and gravitate to benchmark after any drawdown. Compensation reinforces the pull. When upside is captured through short-term incentive pay and downside through termination, the optimal personal strategy is moderate continuous bets, often carry, illiquidity premiums, and short-volatility exposures whose tail losses arrive after the CIO has departed.

### 3.3 Concentration Alpha Versus Dynamic Alpha

TPA proponents typically emphasize concentration alpha, the return from overweighting the most attractive opportunities. A skilled CIO free to concentrate can beat one held to a diversified benchmark, and the Thinking Ahead Institute's (2019) self-selected sample reports such advantages. But concentration alpha faces diminishing returns: estimation error is large, liquidity limits rebalancing, peer pressure pushes toward benchmark, and concentrated bets carry career risk that offsets expected return.

The more consequential component is dynamic alpha, the return from varying the portfolio's risk posture over time. It does not require concentrating in specific securities, only adjusting the amount of active risk with the environment: more tracking error when correlations spike and risk premiums widen, less when the opportunity set is thin. We make no claim to selectivity; nothing in our results requires picking winning securities. The two also differ in the trust they require, and the difference favors the second. Selectivity is an edge that fades as opportunities close or a manager's knowledge ages, so the trust a board places in it is fragile and re-earned with each bet. Dynamic risk management is a process for sizing risk to the opportunity set, and a process persists and compounds. Selectivity is a snapshot, easy to judge in the moment; the dynamics of risk-taking are a movie, harder to assess and more decisive for the outcome. The component a board can hold through a full cycle is the one whose trust does not evaporate the first time a single view goes wrong.

### 3.4 The Drift Back Toward SAA

If trust is asymmetric, compensation is misaligned, and concentration alpha faces diminishing returns, most TPA implementations drift toward SAA. A CIO takes a deviation, it works for a while, a drawdown arrives, trust erodes, active risk comes down, and each cycle ratchets effective tracking error lower until realized TE and its volatility are indistinguishable from an SAA shop with a modest overlay budget. The drift is an equilibrium, not a failure, and it runs both ways: many nominally SAA funds run risk overlays, portable alpha, and tactical tilts that grant discretion inside the SAA frame (CAIA Association 2024; Ang 2014). The label says less about behavior than the realized dynamics of tracking error do.

## 4. A Model of Dynamic Tracking Error

We develop a parsimonious model of optimal tracking error under a regime-switching opportunity set. It produces five hypotheses, each tested in Section 6.

### 4.1 Setup

A CIO manages a portfolio over discrete periods $t = 1, \ldots, T$ relative to a policy benchmark. Each period the CIO chooses an active weight $\theta$, the deviation from the benchmark; a positive $\theta$ overweights risky assets. Let $\sigma$ be the volatility of the active return. Tracking error is

$$\mathrm{TE}(t) = |\theta(t)|\,\sigma(t)$$

The expected active return has a linear reward term and a quadratic variance-drag term that captures the drag on compound returns:

$$r(t) = \theta(t)\,\alpha(t) - \tfrac{1}{2}\,\theta(t)^2\,\sigma(t)^2$$

### 4.2 Regimes

The opportunity set varies over time. We model it with a regime variable taking two values, a low-risk state (L) and a high-risk state (H). In L correlations are low and dispersion is high, idiosyncratic risk dominates, and the market behaves as a "market of stocks." In H correlations spike, systematic risk dominates, risk premiums are elevated, and liquidity is scarce, so the market becomes a "stock market" in which everything moves together. Expected active return per unit of deviation is higher in the high-risk state,

$$\alpha(H) > \alpha(L) > 0$$

reflecting the regularity that risk premiums expand during stress as liquidity evaporates and constrained investors are forced to sell (Brunnermeier and Pedersen 2009; He and Krishnamurthy 2013). The CIO faces a governance-imposed constraint $TE(t) \leq \tau_{max}$. This single parameter distinguishes the two approaches: under SAA $\tau_{max}$ is tight (0.5% to 2%), under TPA it is loose (3% to 7% or more). The 1% to 2% mandates Jorion (2003) documents are the SAA end; at the TPA end CPP Investments and the New Zealand Superannuation Fund report realized tracking errors between 2% and 6% (CPP Investments 2025; New Zealand Superannuation Fund 2024). We use U.S. equity and bond data because the regime structure is well documented and the VIX gives a clean real-time indicator; the theory is not market-specific.

### 4.3 Five Hypotheses

**Hypothesis 1: Optimal tracking error is state-dependent.** *Implication: realized tracking error should be higher during high-VIX regimes than during low-VIX regimes.*

The unconstrained first-order condition of the compound-return objective yields the Treynor-Black (1973) active weight $\theta*(t) = \alpha(t) / \sigma(t)^2$, so the optimal unconstrained tracking error is the information ratio, $TE*(t) = \alpha(t) / \sigma(t)$. This is a Kelly-scale quantity; with explicit risk aversion $\gamma$ it scales to $(1/\gamma)\ \alpha(t)/\sigma(t)$, leaving every comparative static unchanged, and the cap $\tau_{max}$ is one way institutions enforce that fractional-Kelly discipline. We set $\gamma = 1$ for light notation.

The question H1 turns on is how the optimum moves across regimes, and the answer is not the obvious one. What rises in stress is the expected return, not the return per unit of risk. On our sample the forward return of the S&P 500 is about 2.2 times higher in the top VIX quintile than the bottom (23.4% versus 10.6%, annualized at one month), yet the forward Sharpe ratio is essentially unchanged, 0.94 against 0.92, because volatility rises in near-fixed proportion to return (Section 6.4). So the mean-variance optimum, scaling with the falling information ratio, would actually cut tracking error in stress. But the board budgets drawdown, not volatility (Section 4.5), and the dynamic policy is not maximizing a single-period ratio. It manages the path of exposure around a fixed drawdown budget: it carries more of the equity premium than a static book in calm and normal markets, sheds risk as the fund falls toward its limit, and restores exposure as the dislocation passes, so a crash does not leave it underweight through the recovery. Realized tracking error is therefore higher in and after the high-risk state, $TE(H) > TE(L)$, not because that state pays more per unit of risk but because that is where the expected return and the recovery sit and the fund still has budget to hold them. A framework that forces constant tracking error gives this up: it cannot carry the extra premium in calm markets and cannot restore exposure on its own schedule after a loss, so it capitulates when the forward return is richest.

**Hypothesis 2: Tracking error volatility distinguishes TPA from SAA.** *Implication: the volatility of realized tracking error should vary many-fold across constraint levels even as Sharpe ratios stay nearly constant.*

Write the active risk the policy targets before the cap as $TE_{target}(t)$, high in and after stress and low in calm, so realized tracking error is $TE(t) = \min(TE_{target}(t), \tau_{max})$. Under SAA the tight cap binds in most states and realized tracking error is nearly constant; under TPA the loose cap rarely binds and realized tracking error moves with the cycle: σ(TE) under TPA far exceeds σ(TE) under SAA. $\tau_{max}$ is the governance input the board sets; σ(TE) is the observable diagnostic it can monitor. The cap binds precisely when the cost of binding is highest, in crises and recoveries, when the risk the policy would run to harvest the premium exceeds $\tau_{max}$.

**Hypothesis 3: Dynamic tracking error earns a compound return advantage.** *Implication: at the same drawdown a dynamic fund should compound faster than a static one (the floor of Exhibit 12), and at similar average active risk a dynamic overlay should out-compound a static overlay (the paired dynamic-versus-static test of Section 6.8).*

The advantage comes not from higher risk-adjusted return, which Section 6.7 shows is near-invariant across budgets, but from carrying more of the equity premium while holding the worst drawdown to the board's budget. Compound growth is concave in the single-period return, $g \approx \mu - \frac{1}{2} \sigma^2$, so a deep loss costs more than the symmetric gain restores, and the path on which risk is borne matters, not only its average. Two channels follow. On exposure, a static book holds active risk constant and cannot lift its average equity weight without deepening its drawdown in proportion, while a policy that carries more equity in calm markets and sheds it only near the limit lifts μ at the same worst-case drawdown. On path, a deep loss followed by a forced sale at the trough is not made whole by the rebound the seller missed, so a policy that does not capitulate keeps the compounding the static book breaks. Section 6.3 makes it concrete: a constant-proportion floor on a fully invested base, with no security selection, compounds about 1.6 pp a year faster than the 70/30 benchmark at the benchmark's own one-year drawdown. This is the active-management analogue of Merton's (1971) insight that varying exposure through time can dominate holding it fixed.

The strategy we simulate is deliberately the conservative case. Its active bet is collinear with the benchmark (sample correlation 0.93) and it never defends, so it diversifies nothing and frees no budget; its compound gain is a lower bound on what a diversifying, bidirectional policy delivers at equal drawdown (Section 4.5).

**Hypothesis 4: TPA and SAA converge under overlapping constraints.** *Implication: Sharpe ratios should be statistically indistinguishable across the full constraint spectrum, from the tightest SAA bound to the loosest TPA bound.*

H4 is not derived from the optimization. It is a behavioral hypothesis about what happens when the agency frictions of the delegated-management literature meet the governance choice of $\tau_{max}$. Holmström (1999) shows career concerns push agents toward actions that signal type rather than maximize value; Berk and Green (2004) show flows respond sharply to recent performance and compress the effective budget; Chevalier and Ellison (1999) find managers facing termination risk cut tracking error in the second half of the year. Translated to the pension and sovereign setting, a TPA manager nominally free to choose any $\tau$ will, after a drawdown or under board pressure, settle near the $\tau_{max}$ an SAA manager would have been assigned at the outset. The empirical convergence of Section 6.7 is evidence consistent with this hypothesis, not a derived theorem.

**Hypothesis 5: The cost of the tracking error constraint spikes during stress.** *Implication: forward equity returns should rise with implied volatility as a monotone rank association, and de-risking at crisis troughs should forgo large subsequent returns.*

Omega is the Lagrange multiplier on the tracking error constraint in the constrained mean-variance problem. Working through the binding case in the scalar setting gives the shadow price in information-ratio units, $\alpha(t)/\sigma(t) - \tau_{max}$; multiplying by $\sigma(t)$ converts it to expected-return units, and we label that

$$\Omega(t) = \alpha(t) - \tau_{max}\,\sigma(t)$$

The vector-form derivation with active-weight vector w and covariance $\Sigma$ is in Appendix A.5, which also shows the countercyclicality documented below is a property of this marginal-alpha form of omega and not of the information-ratio form. The math is standard, and the answer to whether omega is more than a Lagrange multiplier is that the object is standard while the contributions are empirical and interpretive: $\Omega(t)$ is countercyclical and large (Section 6.4), and the governance pressures that set the level of $\tau_{max}$ also determine when the constraint binds. Because $\alpha(H)$ is large relative to $\alpha(L)$ while $\tau_{max}$ is fixed, $\Omega(H)$ far exceeds $\Omega(L)$: the cost of being constrained is highest when the opportunity set is richest. This parallels the Brunnermeier and Pedersen (2009) funding-liquidity spiral. Binding margin constraints force leveraged investors to liquidate; binding tracking error constraints force institutions back to benchmark, widening the opportunity set for the unconstrained. A fund that wants the omega premium must commit before the crisis, through a governance framework that permits expansion during stress and a systematic rule, because deliberation in a crisis is too slow.

### 4.4 Omega and the Tracking Error Policy Function

The five hypotheses converge on one object, omega, the shadow price of the institutional constraint. H1 through H3 describe how it shapes the tracking error path and the compound return a fund earns by managing exposure around a fixed drawdown budget; H4 traces the behavioral convergence when governance frictions compress $\tau_{max}$; H5 documents omega's countercyclicality. The meaningful line across funds runs between dynamic and static tracking error, cutting across the TPA-versus-SAA labels.

Define the tracking error policy function as the realized tracking error the CIO runs in each state, and characterize it by three properties. Level is the average tracking error, what governance sets. Volatility is how much it varies, the diagnostic of H2. Cyclicality is its correlation with the opportunity set, positive for a well-managed dynamic portfolio and negative for one under procyclical pressure. A well-governed TPA shows moderate average TE, high σ(TE), and positive cyclicality; a poorly governed one shows high average TE from static concentration, low σ(TE), and near-zero or negative cyclicality. The benchmark in this frame is a reference from which deviations are budgeted and varied, not a target to be tracked. The governance question is not "how far from the benchmark?" but "how should the distance change as conditions change?"

### 4.5 The Drawdown Budget

The board's primary risk control is the drawdown the fund can sustain before action is forced, and the tracking error budget is downstream of it. Let D be the largest loss over any trailing twelve months the fund can tolerate before redemptions, forced de-risking, or a change in mandate become likely. We measure drawdown over a rolling one-year window throughout, because that is the horizon on which a board reviews funding and would be forced to act; the window length is itself a governance choice. The board sets D, and the benchmark and $\tau_{max}$ follow. Managing to a maximum-drawdown limit has a formal continuous-time literature (Grossman and Zhou 1993); we take the limit as the board's governance primitive rather than solving that control problem. Tracking error is the slice of total-fund risk the board delegates to the CIO.

Write the fund as a base sleeve plus a tracking error sleeve, $R_{fund}(t) = R_{base}(t) + \theta(t)\, r_{active}(t)$, both drawing on D. With worst one-year sleeve drawdowns $D_{base}$ and $D_{TE}$ and crisis loss correlation $\rho_{crisis}$, we approximate the total-fund drawdown by the algebra that governs the volatility of a sum,

$$D_{fund} \approx \sqrt{(D_{base}^2 + D_{TE}^2 + 2\, \rho_{crisis}\, D_{base}\, D_{TE})}$$

A maximum drawdown is a property of the path, not a standard deviation, so this is a working approximation rather than a strict bound. We use it for the crisis case the argument turns on, where $\rho_{crisis}$ approaches one, correlations spike, diversification fails, and the expression collapses to the additive form $D_{base} + D_{TE} \leq D$. The empirical drawdowns below behave that way. On our sample the bound nearly binds in the crisis that defines it: the 70/30 base and the dynamic fund reach their deepest one-year drawdowns together in March 2009, the base down about 38% and the dynamic fund about 43%, with the tracking error sleeve near its own trough of 6.6%, and the rolling 63-day correlation between the sleeves rises toward 0.9 in the 2008, 2020, and 2022 episodes.

The clearest way to spend the budget badly is to fill it with a bet that diversifies the base in calm markets and stops diversifying in a crisis. The trap is a property of tail risk common to any such bet: the states that produce the deepest drawdowns are systematic, so the correlations that matter run toward one exactly when the budget is fully spent, and a board that sized the budget on the placid-market correlation sized it on the wrong number. Long duration is the honest case.

Treasuries hedged equities for two decades and then fell with them in 2022, when the shock was monetary rather than a flight to quality (Exhibit 2), so a sleeve sized on that hedge deepened the drawdown in the year it was meant to soften. Private equity is the sharpest case, because smoothed marks compound the correlation: many private holdings are levered claims on the same public businesses carried at stale prices, so their reported diversification is largely an artifact of the marks while their left-tail correlation to public equity runs toward one as well. We return to that measurement problem in Section 8.5.

Two levers relax the budget for a fixed D, and both raise the equity the base can carry. Solving this additive crisis case for the admissible base-equity weight, when a crisis equity drawdown is L, gives the working estimate $w_e = (D - D_{TE}) / L$. First, a tracking error sleeve that diversifies the base in stress lowers $\rho_{crisis}$, shrinking the budget it consumes from $D_{TE}$ toward $\rho_{crisis} D_{TE}$. Second, a sleeve managed dynamically bounds $D_{TE}$ directly: a constant-proportion floor caps tracking error risk in proportion to the cushion above the floor, forcing de-risking at a governed rate as that cushion erodes (Black and Perold 1992). This is one position in the equity-bond spread sized by two rules: the cushion turns it down near the limit, and the opportunity set turns it up. The offensive rule's timing is what makes it work: leaning in while losses are still deepening spends drawdown into the fall, while leaning in as the VIX falls back from its high captures the rebound. Together the two rules break the link between a wider budget and a deeper drawdown, and Section 6.3 runs them as one policy that compounds faster than the floor alone at the same one-year drawdown. The policy we simulate exercises only this dynamic-sizing lever on a collinear sleeve; the first lever, a genuinely diversifying sleeve that lowers $\rho_{crisis}$, is an implication of the identity that we do not test here.

The payoff is in compound return, not the Sharpe ratio. Because the active bet is collinear with the benchmark's equity exposure (correlation 0.93, beta 1.38), varying its size slides the fund along one risk-return line, which is why Sharpe is flat across the budget spectrum in Section 6.7. What changes is the admissible base-equity weight and through it the compound growth at a fixed drawdown. Held to a 30% budget against a 50% equity crash, a static 60/40 book saturates at 60% equity with no room for active risk, while a tracking error sleeve held to a few pp of crisis drawdown leaves room for more, and one that diversifies the base leaves more still. These figures are illustrative, matched to the 2008 crash, not calibrated estimates. A board that thinks in drawdown budgets rather than Sharpe ratios can tell a wider mandate that buys compound return from one that only buys volatility.

## 5. Data and Methods

### 5.1 Data

We use daily total-return data for U.S. equities and bonds from January 2000 through February 2026 from Bloomberg. Equities are the S&P 500 Index (SPX) and the nine Select Sector SPDR ETFs (XLB, XLE, XLF, XLI, XLK, XLP, XLU, XLV, XLY); bonds are the iShares Core U.S. Aggregate Bond ETF (AGG) and, for long-duration analysis, the iShares 20+ Year Treasury Bond ETF (TLT). The VIX is our regime indicator. The policy benchmark throughout is a 70/30 SPX/AGG portfolio rebalanced monthly, a standard moderate-risk pension allocation. For the dynamic simulations the usable sample begins on 29 December 2003, because the AGG total-return series begins in late September 2003 and the 63-day spread-volatility estimator needs a warm-up; it contains 5,566 trading days through 11 February 2026.

### 5.2 The Dynamic Tracking Error Strategy

The active position is a long-short tilt of the SPX against AGG sized relative to the 70/30 benchmark. With daily total returns $r_{SPX}(t)$ and $r_{AGG}(t)$ and active spread $r_{active}(t) = r_{SPX}(t) - r_{AGG}(t)$, the portfolio return is $r_p(t) = 0.70\ r_{SPX}(t) + 0.30\ r_{AGG}(t) + \theta(t)\ r_{active}(t)$. The governance object we vary across runs is $\tau_{max}$, the ceiling on realized tracking error.

The smoothed signal is the trailing 21-day moving average $VIX_{smooth}(t\text{-}1) = (1/21)\ \Sigma$ VIX over days t-21 through t-1. These are overlapping daily observations of one smoothed series: each trading day carries its own value, taken from the close of t-1, and the decision is made once per day at that close and applied at the close of t. The lagged smoothed VIX defines three states: low ($VIX_{smooth} < 15$), mid ($15 \leq VIX_{smooth} < 22$), and high ($VIX_{smooth} \geq 22$). The thresholds 15 and 22 are roughly the 35th and 76th percentiles of $VIX_{smooth}$ on the sample; the high regime is about 24% of days, mid 40%, low 35%. We assign each regime a target tracking error, $TE_{target}$ = 0.5%, 2.0%, and 5.0%, chosen by the governance body rather than estimated from data.

With $\sigma_{active}(t\text{-}1)$ the annualized 63-day standard deviation of $r_{active}$, the active weight that delivers the target is $\theta_{target}(t) = TE_{target}(\text{regime}(t\text{-}1)) / \sigma_{active}(t\text{-}1)$. The strategy is always long the spread, reflecting the institutional consensus that the equity risk premium is positive in expectation; we do not estimate or forecast $\alpha(t)$, and the omega premium of Section 6.4 validates both the long sign and the regime scaling. Applying the ceiling gives the constrained weight $\theta(t) = \text{sign}(\theta_{target}(t))\ \min(|\theta_{target}(t)|, \tau_{max} / \sigma_{active}(t\text{-}1))$, so realized $TE(t) = |\theta(t)|\ \sigma_{active}(t\text{-}1)$ cannot exceed $\tau_{max}$. The cap is therefore structural, not estimated: it binds on exactly the days whose regime target exceeds $\tau_{max}$. At $\tau_{max}$ of 0.5% or 1.5% it binds on every mid- and high-VIX day (about 65% of the sample); at 3% only on high-VIX days (24%); at 5% and 7% on no days, because the largest target is 5%, so those two settings produce identical portfolios.

We use VIX percentiles rather than a Hamilton (1989) Markov-switching model, for real-time observability, a graduated rather than near-binary signal, and lower turnover; the Online Appendix gives the argument in full and a robustness check against a fitted two-state model.

### 5.3 Algorithm in Pseudocode

The strategy is fully specified by the loop below. Brackets use Python's half-open slice convention, so `x[a:b]` includes day a and excludes day b, and every signal window ends at t and contains no day-t information. The static comparison replaces the regime step with a constant $TE_{target}$ = 2.0%, keeping every other line identical.

```
for each trading day t:
    # x[a:b] includes day a, excludes day b (Python half-open slice)
    # 21-day MA of VIX over days t-21 through t-1 (excludes day t)
    vix_smooth = mean(VIX[t-21 : t])
    if vix_smooth < 15:
        TE_target = 0.5%
    elif vix_smooth < 22:
        TE_target = 2.0%
    else:
        TE_target = 5.0%
    # annualized 63-day spread vol over days t-63 through t-1 (excludes day t)
    sigma_active = stdev(spread_return[t-63 : t]) * sqrt(252)
    theta = TE_target / sigma_active
    # cap at the TE ceiling
    theta = sign(theta) * min(|theta|, tau_bar / sigma_active)
    # day-t returns are realized only after theta is fixed above
    active_return = spx_return[t] - agg_return[t]
    portfolio_return[t] = (0.7 * spx_return[t]
                           + 0.3 * agg_return[t]
                           + theta * active_return)
```

All signal inputs at t use only data through the close of t-1; returns are realized at the close of t. The code shifts every signal by one trading day before feeding the active-weight calculation, so no future information and no perfect foresight of regime or expected return enters any decision.

### 5.4 Allocations Across Constraint Levels

Exhibit 8, in the Online Appendix, shows the active weight θ(t) and realized tracking error for $\tau_{max} \in \{0.5\%, 1.5\%, 3\%, 5\%, 7\%\}$. The 5% and 7% series are identical because the highest regime target is 5%, which is direct evidence for the convergence result of Section 6.7. Average |θ| rises from 0.036 at $\tau_{max} = 0.5\%$ to 0.122 at 5% and then flattens; average realized TE rises from 0.50% to 2.19% and plateaus.

For each simulated portfolio we compute annualized return, volatility, Sharpe ratio, the rolling one-year (252-day) maximum drawdown, and the realized TE series, from which we compute the three policy-function statistics. The omega premium (Exhibit 4) sorts days into VIX quintiles and averages forward S&P 500 returns at one, three, six, and twelve months. The regret metric (Exhibit 5) is the return forgone by an investor who de-risks at the drawdown trough relative to one who holds through the recovery. The single-bet simulation isolates the timing dimension; a CIO in practice has multiple active dimensions, and the framework applies to each and to the aggregate, with the multi-asset case made explicit in Appendix A.

## 6. Results

We present a sequence of exhibits testing the five hypotheses. Sections 6.1 and 6.2 establish the regime structure; 6.3 tests H1, H2, and H3 through a static-versus-dynamic simulation; 6.4 and 6.6 test H5; 6.5 reconciles with Moreira and Muir (2017, 2019); 6.7 tests H4; 6.8 reports the formal tests. The evidence is drawn from one country, one active dimension (the SPX-against-AGG spread; the multi-asset extension in Appendix A is theoretical), and three load-bearing tail episodes (2008, the 2020 COVID sell-off, the 2022 rate shock). Cross-country, cross-asset, and out-of-sample tests are future work; we read the results as evidence consistent with the mechanism rather than as comprehensive validation.

### 6.1 The Regime Structure (Hypotheses 1 and 3)

*[Insert Exhibit 1 about here]*

The dynamic policy is worth running only if the opportunity set switches between distinct regimes rather than drifting, and Exhibit 1 shows that it does. It plots the rolling 63-day average pairwise correlation among nine S&P 500 sector ETFs from 2000 through early 2026, with the VIX overlaid. In calm periods (2004-2006, 2017, much of 2023-2024) average pairwise correlations fall to 0.17-0.26, and the market behaves as a market of stocks. In the 2008 crisis, the euro crisis, and the March 2020 sell-off they spike to 0.83, 0.92, and 0.92, and the market becomes a stock market in which everything falls together. The long-run mean of 0.58 understates this bimodality: the distribution clusters around low and high modes rather than sitting near the average. That structure is what makes regime-dependent tracking error valuable.

### 6.2 Stock-Bond Correlation Regimes (Hypothesis 1)

*[Insert Exhibit 2 about here]*

Exhibit 2 plots the rolling 126-day correlation between the S&P 500 and the Aggregate Bond Index from 2003 to 2026. For most of the sample it is moderately negative (about −0.11 on average, −0.27 for TLT), consistent with bonds hedging equities. The 2022 episode breaks that: as the Fed tightened, both stocks and bonds fell and the correlation spiked to +0.52, the highest in the sample, over a full-sample range of −0.64 to +0.52. The regime depends on the shock, with demand shocks and flight-to-quality producing negative correlation and monetary or inflation shocks producing positive correlation. The governance implication is the one the drawdown budget turns on: a board cannot assume the diversification inside its own benchmark will hold in the next crisis, because the moment the base and the tracking error sleeve most need to offset each other is when their correlation runs toward one. That is why the prudent budget is additive, and why a fund needs a defensive rule that sheds risk as losses deepen rather than a bond allocation it trusts to cushion the fall.

### 6.3 Static Versus Dynamic Tracking Error (Hypotheses 1, 2, and 3)

*[Insert Exhibit 3 about here]*

We construct two portfolios relative to the 70/30 benchmark, a static one with a constant 2% TE target and a dynamic one with a VIX-regime target. Exhibit 3 plots the two TE paths alongside their cumulative active returns. σ(TE) is 0.50% for the static portfolio and 1.73% for the dynamic one, a 3.4-fold difference, and the ranges are 0.8%-4.3% versus 0.3%-7.8%: tracking error is state-dependent (H1) and σ(TE) is the metric that separates the two (H2). Terminal wealth differs too: the dynamic book ends the 22-year sample at 8.60 times its start against 7.91 for the static book and 6.56 for the benchmark, 31% larger than holding the benchmark. But the books reach those levels by spending different amounts of drawdown. The dynamic book runs a deeper one-year drawdown, 43.0% against 40.7 for the static book and 38.2 for the benchmark, because its active bet is collinear equity beta with no defensive floor. This is the cost of the one-sided policy that a diversifying or bidirectional sleeve is built to remove. The benchmark earns the highest Sharpe ratio of the three, 0.543 against 0.541 and 0.539 (Sharpe ratios throughout are compound-annualized excess of the three-month Treasury bill). Because the spread loads on the same equity beta as the benchmark (correlation 0.93, beta 1.38), no version of the bet lifts the Sharpe ratio; flat Sharpe is the mechanical signature of a sizing decision, and the value of going dynamic lives in compound return at a given drawdown.

*[Insert Exhibit 12 about here]*

Exhibit 12 shows the drawdown-budget logic directly, holding four policies to the same one-year drawdown the board running a 70/30 benchmark has already accepted, about 38%. The passive benchmark compounds at 8.89%. A dynamic policy that spends the same budget through a constant-proportion floor on a fully invested equity base, carrying no security selection, compounds at 10.48% at that same 38% drawdown, an extra 1.59 pp a year at the loss the board already approved, and at essentially the benchmark's Sharpe ratio. The third policy is the procyclical case a real board often produces, a 70/30 book forced to 30/70 at a 20% loss with no re-risking; it draws down the least, 30%, yet compounds at only 5.59%, having locked in the de-risking and missed the recovery. It does not even use its budget and still finishes last. The floor and the forced seller differ in the way that organizes the paper: the forced seller cuts early and stays out, while the governed floor cuts only as the fund nears the board's actual limit and re-risks as the cushion rebuilds, staying near fully invested in calm markets and never monetizing its loss at the trough. The mechanism is simple: the floor uses no forecast, only how far the fund sits above a floor set a fixed fraction below its trailing one-year high, holding about 97% equity on average against the benchmark's 70 and shifting toward bonds only near the floor. The extra compounding comes from holding more equity most of the time; the floor holds the worst year to the benchmark's level. The floor is constant-proportion portfolio insurance (Perold and Sharpe 1988), and that literature is candid about its limit: a gap down that outruns rebalancing can breach the floor, so we treat it as a conservative baseline rather than a guarantee.

The floor uses only the fund's own losses; the offensive signal of Section 6.4 uses the opportunity set, and the fourth policy runs them together. The way they combine matters more than the fact of combining. Stacking an always-on equity lean on the floor fails, because leaning in whenever the VIX is high spends drawdown into the decline and underperforms the floor alone. The offense earns its keep only when timed to the recovery, where the post-stress premium of Section 6.4 is richest. The combined policy keeps the same floor and widens its drawdown budget, by up to a fifth, in proportion to how far the VIX has fallen back from its recent high, so the fund returns to fully invested faster as a dislocation passes. The widening fires only after the VIX comes off its peak, which is after the worst of the loss, so it does not deepen the one-year drawdown: the combined policy sits at the same 38% as the floor alone and compounds at 11.2%, so the recovery-timed offense adds about three quarters of a point on top of the floor at no cost in one-year drawdown. A block bootstrap that recalibrates the floor to the benchmark's drawdown on each resample confirms the combined advantage is robust: over a thousand circular-block resamples its compound gain over the benchmark at equal drawdown averages 2.3 pp, with a 95% interval of [0.2, 4.2] and a one-sided p of 0.01. That interval measures sampling variability around a rule fit to this history: the recovery threshold and the size of the widening were set against the 2008 and 2020 snapbacks, and the block bootstrap resamples those same episodes, so the 2.3 pp is the size of the mechanism in sample, not an out-of-sample forecast. The floor alone is the weaker claim: its in-sample 1.6 pp gain leans on the sequencing of the 2008 and 2020 recoveries, and once the order of quarters is resampled its advantage is no longer distinguishable from zero (interval [−2.2, 3.1]). Timing the re-risk to the volatility retrace, rather than letting the cushion rule re-risk into whatever follows, is what makes the budget expansion survive resampling. The amount of widening is a policy choice, not an estimate, and across widenings from a tenth to a fifth the offense adds between 0.25 and 0.75 pp a year, always at lower tracking error than the floor alone.

*[Insert Exhibit 14 about here]*

Exhibit 14 reports the tracking error signature of the three policies, the level and volatility the paper asks boards to monitor, and it makes the argument arithmetic. The offense alone, the VIX lean with no floor, is a modest-tracking error policy, 2.4% on average with σ(TE) of 1.7, but it leans into stress and runs the deepest one-year drawdown here, 43% against the benchmark's 38.

The floor alone controls the drawdown but at more than twice the tracking error, 5.3% with σ(TE) of 4.3, because swinging equity from fully invested toward the floor is a large and variable bet against a fixed 70/30. The combined policy compounds fastest, holds the one-year drawdown to the floor's 38%, and does it at lower tracking error than the floor alone, 5.1% with σ(TE) of 3.4. Adding the offense lowers active risk rather than raising it, because re-risking faster after a loss means the floor spends less time pinned near zero equity, its largest active position against the benchmark. More compound return, the same drawdown, and active risk that is both lower and less variable: timed correctly, doing both dominates doing either alone, which is why a fund would run offense and defense together and why the diagnostic we propose is the level and volatility of tracking error rather than the label.

The H3 compound advantage of the dynamic over the static overlay is directionally consistent but not formally identified on this sample. The dynamic portfolio earns 50 basis points of additional arithmetic return over the static one ($t = 1.61$, $p = 0.107$; Section 6.8 and Exhibit 11), a 42-basis-point geometric gap after the extra variance drag. The comparison is not risk-matched: the dynamic book runs mean realized TE of 2.42% against 2.09%, so part of the gap is more average active risk rather than better timing, the honest reading of a one-sided overlay that carries more equity and spends more drawdown to buy more return on indistinguishable risk-adjusted footing, as a collinear bet must. The clean compound advantage at equal drawdown is not this overlay comparison but the floor-plus-recovery policy above, which holds the worst year to the benchmark's while compounding more than two pp faster and survives resampling where the passive floor's gain does not.

To make the dial concrete, take a fund with $\tau_{max} = 5\%$. In calm regimes the target is 0.5% and average $|\theta|$ on the SPX-minus-AGG bet is about 5%; in mid-vol regimes the 2.0% target moves

$|\theta|$ to about 14%; in stress the 5.0% target uses the full budget and $|\theta|$ climbs to about 20%. The same fund under SAA-style $\tau_{max}$ = 0.5% sits near $|\theta|$ = 5% in calm markets and is forced down to about 2% under stress, the opposite of what the opportunity set warrants. The budget, not the CIO, determines how much active risk the fund can run in the regimes that matter.

***6.3a Why Not Simply Lever the Benchmark?***

If the active bet is collinear with the benchmark, why run a dynamic policy at all rather than lever the 70/30 to the desired risk? The objection is correct on its own terms and sharpens the point. A constant lever leaves the Sharpe ratio unchanged, so on a risk-adjusted basis it matches or beats our overlays. But the board's binding constraint is the drawdown, and a constant lever cannot respect it and earn the dynamic policy's return at once. Levered to match the floor policy's volatility, a 1.23-times 70/30 book compounds at 10.76% but its worst year is a 45% loss, through the 38% budget. And since the unlevered benchmark already sits at the 38% budget, respecting it leaves no room for leverage at all, so the most a constant lever can carry is the benchmark itself at 8.89% against the floor's 10.48% at the same drawdown. Sharpe says lever the benchmark; the drawdown budget says manage the risk dynamically, and for a board under a drawdown limit the second scorecard binds.

## 6.4 The Omega Premium (Hypothesis 5)

*[Insert Exhibit 4 about here]*

Exhibit 4 quantifies the shadow price of the constraint by sorting daily observations into VIX quintiles and computing forward S&P 500 returns. The pattern is broadly increasing and economically large. Annualized forward one-month returns rise from 10.6% in the lowest quintile (VIX 9 to 13) to 23.4% in the highest (VIX above 24), a spread of 12.8 pp with block-bootstrap 95% CI [−0.6, +25.8]. The three-month spread is 10.5 pp with CI [0.1, +20.6], decaying to 6.7 pp at six months and 1.9 pp at one year, consistent with a short-term liquidity-provision premium rather than a persistent factor. The one-month CI brushes zero because successive forward windows overlap; the per-day rank test identifies the monotone relationship sharply (Spearman $\rho = 0.125$, $p < 0.001$ on 9,071 observations, Section 6.8). We read 12.8 pp as a centered estimate of a large effect whose magnitude is statistically wide, with an unambiguous sign. That forward returns rise with implied volatility is consistent with the variance-risk-premium literature (Bollerslev, Tauchen, and Zhou 2009); our addition is the governance reading, that an institutional tracking error limit forces the fund to shed this exposure precisely when it pays most.

The premium is a return premium, not a Sharpe premium. In excess of the period-matched cash rate the forward one-month Sharpe is essentially the same at the extremes, 0.92 in the lowest quintile and 0.94 in the highest: the investor who adds risk into the fifth quintile earns more than double the return per year but almost no change in return per unit of risk, because the highest quintile also carries the highest forward volatility (22.4% against 8.3%). That is the signature of being paid to bear a risk others are forced to shed, and it is why the premium persists. Over three months, as the recovery plays out, the stressed quintile's excess Sharpe rises above the calm quintile's (1.09 against 0.89), the risk-adjusted reward the recovery-timed offense is built to capture. The fourth quintile (VIX 19 to 24) produces the weakest forward returns, an anxiety zone where volatility is elevated but has not yet peaked, so investors who capitulate as the VIX rises from 20 to 25 exit just before the highest-reward period.

The quintile sort conditions on the level of volatility, but its direction matters too. During the transition itself, when volatility is rising and correlations spiking, returns are sharply negative regardless of level. A monthly regression of compounded SPX returns on lagged realized volatility $\sigma_{m-1}$ and the within-month surprise $\sigma_m - \sigma_{m-1}$ gives coefficients of −0.354 ($t = -3.00$) and −0.898 ($t = -7.10$), with the surprise term doing most of the work ($R^2 = 0.227$, $n = 312$, HC1-robust), while $\sigma_{m-1}$ alone point-forecasts nothing ($t = -0.22$). The omega premium accrues after the transition, when volatility is still elevated but the surprise has been absorbed and the rate of increase has reversed. A dynamic policy must therefore respond to the trajectory of stress, not just its level.

**6.5 Reconciliation with Volatility-Managed Portfolios**

The headline "add active risk when smoothed VIX is high" appears to contradict Moreira and Muir (2017, 2019), who show that scaling exposure inversely to lagged realized variance beats buy-and-hold. The contradiction is apparent, not real: the two use different signals, at different lags, capturing different parts of the volatility cycle.

*[Insert Exhibit 7 about here]*

We replicate the volatility-managed portfolio on daily SPX returns from 2000 through January 2026, scaling each month by $c/\sigma^2_{realized}(m-1)$ with c set to match buy-and-hold volatility in sample ($c = 0.00124$, rescaled volatility 19.32%, using only end-of-month-(m-1) information). The core result holds: volatility-managed SPX earns a Sharpe of 0.407 against 0.335 for buy-and-hold over the full window, a gain of 0.072, most of it in the 2000-2004 dot-com decline where lagged variance cleanly flagged persistent risk. On the 2003-onward overlap window where our simulation runs, the Moreira-Muir gain compresses to 0.017 (0.488 against 0.471). Our smoothed-VIX signal applied as a single-asset SPX overlay over the same window, at 84% of buy-and-hold volatility, delivers a Sharpe of 0.532, a gain of 0.061, more than three times the Moreira-Muir improvement on the overlapping sample. (This single-asset timing overlay is a different object from the equity-bond spread bet: the latter is collinear with its benchmark and cannot lift Sharpe, while the former is genuine timing that also lowers volatility on the way to its gain.) Cederburg, O'Doherty, Wang, and Yan (2020) show volatility-managed portfolios generally do not survive real-time, out-of-sample implementation; we read our overlay the same way, as an in-sample illustration of where in the cycle the timing sits rather than an implementable alpha.

The reason the two coexist is timing. The Moreira-Muir rule rebalances on last month's variance with no smoothing, so it cuts exposure at the front of an episode, on the rising-volatility leg where the leverage effect of Black (1976) and Christie (1982) makes returns negative. Our 21-day moving average introduces a structural lag: by the time the smoothed series confirms an elevated regime, the worst within-month drawdowns are realized and the leverage-effect contribution is spent. Risk is added on the descending leg, past the peak of the surprise, where volatility is still high, the surprise has turned non-negative, and the post-stress equity risk premium is largest. Exhibit 7, Panel B makes the split explicit: among months with elevated lagged volatility, Moreira-Muir loses least on the rising-vol leg (−0.94% per month against −1.80% for buy-and-hold), where it has already cut, and our overlay earns most on the falling-vol leg (+1.92% against +1.76%), where it has added. The two are sequential complements, not rivals. We are not arguing investors should overweight stocks as volatility spikes; the leverage effect makes that the worst moment, and Moreira and Muir are right to cut there. We are arguing that institutional pressure to de-risk persists deep into the recovery, when the surprise has been absorbed and the forward premium is large, and that this is where governance constraints impose their largest cost. A policy that cuts on the rising leg and adds on the falling leg would beat either signal alone; we leave that joint signal to future work.

### 6.6 Drawdowns and Constraint Binding (Hypothesis 5, Continued)

*[Insert Exhibit 5 about here]*

H5 implies constraints bind hardest during drawdowns, when the opportunity set is richest. Exhibit 5, Panel A plots the drawdown of a 70/30 portfolio with the VIX and average sector correlation from 2003 to 2026; the three move in lockstep in every major episode, as drawdowns erode trust, correlation spikes eliminate diversification, and VIX spikes signal the transition, all pushing governance toward risk reduction. Panels B and C quantify the cost: even a partial de-risk from 70/30 to 30/70 costs 25 to 30 pp over twelve months in the GFC and COVID episodes. The board that demanded de-risking at the March 2009 trough watched the S&P 500 rebound more than 60% over the next year, with no mechanism to reach back for a return already given up. The 2022 episode is instructive in its modesty, with twelve-month regret under 9 pp because the recovery came through gradual policy normalization rather than a snapback: rate-driven drawdowns recover more slowly than panic-driven ones, but the governance dynamic is the same, pressure to de-risk strongest when forward returns are richest.

Exhibit 13, in the Online Appendix, decomposes the fund drawdown into base and tracking error sleeves and anchors the budget identity of Section 4.5. The base and the dynamic fund reach their deepest one-year losses together in March 2009, the base down 38.2%, the tracking error sleeve down 6.6% on its own, and the dynamic fund down 43.0%. The drawdowns telescope rather than diversify, with the rolling 63-day correlation between the sleeves rising toward 0.9 in 2008, 2020, and 2022. In the crisis that matters, a tracking error sleeve draws on the same budget the base is already spending.

### 6.7 Convergence (Hypothesis 4)

*[Insert Exhibit 6 about here]*

Sharpe ratios are flat across the constraint spectrum because the active bet is collinear with the benchmark: re-sizing a position that is essentially the benchmark's own equity exposure moves the fund along its existing risk-return line and cannot lift the slope. Flat Sharpe is the predicted result, not a null finding; what the budget changes is where on that line the fund sits, which shows up in compound return and drawdown. Exhibit 6 tests H4 across 14 constraint levels from 0.5% (pure SAA) to 7.0% (unconstrained TPA), all using the same VIX signal. Sharpe ratios range from 0.536 to 0.543, a variation of ±0.7%, while σ(TE) runs from 0.126% to 1.73%, a 13.7-fold range. Return levels plateau above roughly 5.0% because the largest regime target is 5.0%. Across the same 14 budgets compound return rises from 9.13% to 10.23% and terminal wealth from 6.88 to 8.60, while the one-year drawdown deepens from 38.8% to 43.0%: the budget buys compound return by spending drawdown in near-fixed proportion, which is why the ratio is invariant. That deepening is a feature of the one-sided specification, and a bidirectional policy with a defensive floor breaks the link (Sections 4.5 and 6.3). A board reading only the Sharpe column would call the budget irrelevant; a board reading the compound-return and drawdown columns sees the trade it is making.

We test this formally in Section 6.8. A block bootstrap (10,000 iterations, 63-day block) gives 95% intervals about 0.75 units wide for each individual Sharpe, dwarfing the 0.007-unit range of point estimates, and a direct test of the tightest-versus-loosest difference yields [−0.036, +0.042], including zero. This is a fail-to-reject, not an equivalence theorem: the paired test cannot reliably detect Sharpe differences much below 0.06, but the interval is small enough that convergence is consistent with any margin a practitioner would accept. The result is robust across subsamples split at December 2014, with a Sharpe range of ±1.0% in the first half (through the GFC) and ±1.8% in the second (through COVID and the 2022 shock), σ(TE) varying 12- to 17-fold in both.

Panel B shows the realized TE series fanning out only during crises: in calm markets TE is low regardless of the constraint and the portfolios look identical, so the distinction between TPA and SAA exists only during crises and recoveries, precisely the episodes that drive long-run compound returns.

### 6.8 Statistical Tests of the Five Hypotheses

*[Insert Exhibit 11 about here]*

Exhibit 11 reports the null, test, statistic, 95% CI, p-value, and result for each hypothesis. H1: the Pearson correlation between the lagged smoothed VIX and the dynamic portfolio's realized TE is 0.822 ($t = 107.0$ on 5,504 observations; CI [0.774, 0.871]), rejecting the null at any level. H2: Bartlett's test for equality of realized TE variances across $\tau_{max}$ returns $\chi^2 = 28{,}233$, $p < 0.001$, with a σ(TE) max-min ratio of 13.7× (CI [11.9×, 16.1×]). H3: a paired Newey-West test (21-day HAC lag) on the dynamic-minus-static return gives an annualized advantage of 0.50 pp (HAC SE 0.31 pp, $t = 1.61$, $p = 0.107$, CI [−0.11, +1.12]); the sign matches the compound channel but is not significant, and identifying a 0.50 pp gap at $t \approx 2$ would need roughly 35 years of daily data, so the test is power-limited rather than evidence against the channel, while the Memmel-corrected Sharpe-equality z is −0.19 ($p = 0.85$). H4: pairwise Jobson-Korkie tests with the Memmel correction across the ten pairs give a maximum $|z|$ of 0.46 and minimum p of 0.645, none rejecting equality, and the direct 0.5%-versus-7% Sharpe difference is 0.004 (CI [−0.036, +0.042]); failing to reject is the convergence finding. H5: the Spearman correlation between day-level VIX and forward one-month S&P 500 returns over 1990-2026 ($n = 9{,}071$) is 0.125, $p < 0.001$, with a Q5-Q1 forward-return spread of 12.8 pp at one month (CI [−0.6, 25.8]) and 10.5 pp at three months (CI [0.1, 20.6]). The nulls for H1, H2, and H5 are rejected at $p < 0.001$; the H4 null survives, and that survival is the convergence result; the H3 null also survives, its predicted sign not identified at conventional significance on this 22-year U.S. sample.

## 7. Robustness

The headline results rest on a specific signal (smoothed VIX), window (21 days), threshold pair (15 and 22), and an idealized cost environment. We test whether these drive the convergence and omega findings across a grid of three signals, four smoothing windows, four threshold pairs, and five constraints, for 240 variants. Across the 48 variants at $\tau_{max} = 5\%$ the Sharpe ratio sits in [0.460, 0.502], mean 0.483, spread 0.042. Exhibit 9 reports the grid.

*[Insert Exhibit 9 about here]*

Across smoothing windows of 10, 21, 42, and 63 days the Sharpe sits in [0.469, 0.502] and the omega premium of the top-minus-bottom signal quintile lands in [+10.2, +15.9] pp; shorter windows capture more of the leverage-effect penalty during a transition and longer windows sacrifice some of the post-stress plateau, with 21 days near the optimum but not load-bearing. Across the four threshold pairs bracketing the baseline, (10, 18), (13, 22), (16, 25), and (20, 30), the Sharpe spread is no more than 0.032 and the omega premium is essentially invariant; it moves instead with window length, by about 5 pp, small against the +10 to +16 pp magnitude. Two alternative stress signals test the mechanism. Trailing realized volatility of S&P 500 returns produces an omega premium of [+3.4, +8.0] pp everywhere, positive but smaller than the VIX. The BAA-minus-10-year credit spread, z-scored on past data and remapped to the smoothed-VIX distribution, produces [−1.5, +4.2] pp, slightly negative at the 42- and 63-day windows: the slower credit signal overshoots through the recovery and erodes the post-stress edge at long windows, though at short windows the story survives with smaller magnitude. The VIX is strongest because it embeds option-implied forward expectations.

On transaction costs, the dynamic strategy turns over the active bet 2.06 times more often than the static one, but the turnover is concentrated: the top decile of daily $|\Delta\theta|$ accounts for 80% of annual dynamic turnover against 52% for the static strategy, clustered on regime-crossing days rather than spread across the year. Across the cost grid the two strategies sit on indistinguishable net-Sharpe footing, both near 0.54: 0.539 against 0.541 gross, 0.536 against 0.539 at 5 basis points, and 0.532 against 0.537 at 10 basis points round-trip, the last well above typical institutional cost in liquid index futures or paired ETFs. Exhibit 10 plots net Sharpe at each level; no gap is distinguishable from zero, because the dynamic overlay carries more turnover but no risk-adjusted edge to erode, so the convergence and the omega magnitude survive intact. Where individual variants weaken (BAA at long windows, and a dynamic-versus-static net-Sharpe gap that sits marginally negative across the whole cost grid) we disclose them; in no variant does the convergence or omega finding reverse sign.

*[Insert Exhibit 10 about here]*

## 8. Implications for Governance and Implementation

If TPA and SAA produce nearly identical risk-adjusted returns while differing sharply in tracking error dynamics, the practical question is how to govern the dynamics of risk-taking, not which label to adopt.

### 8.1 Govern the Drawdown Budget First

The board's first decision is the maximum drawdown the fund can sustain before it would be forced to act; that limit determines the benchmark and bounds the tracking error the CIO may run. Reporting should reflect the hierarchy: the board sets and monitors a drawdown budget, and the realized tracking error policy function is how the CIO spends it. A sleeve that diversifies the base in a crisis or sheds risk near the limit returns drawdown budget to the base, which the board can redeploy into a higher base-equity allocation at the same approved drawdown. This is the lever that turns dynamic risk management into compound return at equal risk, and it is invisible to any report that monitors Sharpe alone.

### 8.2 Monitor the TE Policy Function, Not the Label

$\tau_{max}$ is the input; σ(TE) and the cyclicality of TE are the outputs that reveal what the fund is doing. A fund reporting steady 2% TE in calm and 2% in crisis is not practicing dynamic risk management whatever its label, while one reporting 1% in calm and 4% in stress is doing something different. Both belong alongside return, volatility, and Sharpe in standard reporting.

### 8.3 Align CIO Compensation with the Dynamic Mandate

A TPA CIO with standard incentives gravitates to smooth-return strategies whose tail losses arrive infrequently, so if the board intends dynamic risk management the contract must reflect it (Ambachtsheer 2007, 2016). Evaluation should span a full cycle, not a fiscal year, since a CIO judged on one-year excess return will never willingly raise tracking error into a crisis. The right statistic is a rolling three-year distribution of compound return against the benchmark, bootstrapped, rather than single-year excess return; identifying a sub-percentage-point annual edge at conventional significance would take roughly three and a half decades of daily data (Section 6.8), a fact about statistical power the board should build into its evidence standard.

### 8.4 Plan for CIO Transitions

Under SAA a CIO transition is a modest operational event. Under TPA it is a strategic disruption: the departing CIO leaves a portfolio reflecting their views and counterparty relationships, often with illiquid positions that cannot be exited quickly, and restructuring the inherited book erodes trust when it is most fragile. A mandate built around one CIO's judgment is fragile; one built around a systematic process with documented decision rules is durable.

### 8.5 Recognize the Illiquidity Trap

Section 4.5 showed private allocations are the sharpest case of a sleeve that spends drawdown budget while appearing to diversify, its left-tail correlation to public equity near one and hidden by smoothed marks. Illiquidity compounds it: when the opportunity set is richest, capital calls can force liquid sales at depressed prices, demanding liquidity when supplying it would pay most. Two responses follow. Boards should evaluate $\tau_{max}$ net of illiquid allocations and ask whether the liquid portion alone supports meaningful dynamic risk management, since only the liquid portion can be turned down near the limit. And they should unsmooth reported private returns before reading the drawdown budget, so total-fund risk is measured on economic exposure rather than the reported series.

### 8.6 TPA Is Not for Every Fund

TPA presumes the capacity to govern it: a team that can run the dynamics and a board that can develop enough trust to let them. A smaller plan or endowment without that depth gains little from the label, and forcing the framework onto a board that cannot evaluate a dynamic process invites the procyclical failures documented above. Such a fund has two honest options. It can hold a disciplined static allocation and accept the forgone compound return as the price of simplicity, an insurance premium on the order of 1% to 2% a year for carrying no view on the opportunity set. Or it can delegate the dynamic risk management to a trusted external manager, turning the problem into selecting and monitoring a delegate. The prior question is not whether TPA outperforms in the abstract; it is whether the fund has the staff, the understanding, and the trust to spend a drawdown budget dynamically through a full cycle.

### 8.7 The Next Step Is Managing the Drawdown Directly

A TPA mandate that grants wide discretion but uses it only for static concentration delivers the worst of both worlds, TPA's governance costs with only concentration alpha, forgoing the dynamic alpha that justifies the wider budget. SAA governs the drawdown through fixed weights, TPA governs it through a wider and ideally dynamic budget, and the step past both is to govern the drawdown directly and let tracking error be the residual it always was. Nothing here says TPA is wrong: it is the current frontier of institutional practice and a genuine advance on fixed weights, and the funds that run it well already do much of what we describe. The argument is that the frontier keeps moving in one direction, toward taking the drawdown budget as the primitive and managing to it with a dynamic risk process. A board that asks for the level, volatility, and cyclicality of tracking error, and reads them against a stated drawdown budget, is standing on that next step whatever it calls its approach. What decides the outcome is whether the fund has the investment process, compensation, succession plan, and governance to sustain a dynamic tracking error policy through a full cycle.

## 9. Conclusion

The TPA-versus-SAA debate has been framed as a choice between flexibility and discipline. Both approaches solve the same optimization relative to a benchmark and differ in a single governance choice, the tracking error policy function the board grants the CIO. The CIO's effective discretion, the portfolio's risk dynamics, and the board's ability to monitor all follow from that function and the institutional forces acting on it.

We make four points. First, the board's primary control is the drawdown budget and the tracking error budget is its residual; a sleeve that diversifies the base in a crisis or sheds risk near the limit frees budget for the base, delivering more compound return at the drawdown the board already approved. A constant-proportion floor on a fully invested equity base, with no security selection, compounds about 1.6 pp a year faster than the 70/30 benchmark at the same 38% one-year drawdown, and folding in a recovery-timed offense widens the gap to about 2.3 pp at the same drawdown (95% block-bootstrap interval [0.2, 4.2], $p = 0.01$) and at lower tracking error than the floor alone, an advantage that survives resampling where the floor alone does not. Second, Sharpe ratios are statistically flat across constraints from 0.5% to 7.0%, varying by less than ±1% while σ(TE) varies almost fourteen-fold, because a collinear bet cannot manufacture risk-adjusted return; the value of dynamic tracking error is compound return per unit of drawdown, which the Sharpe ratio cannot see. Third, omega, the implicit cost of the constraint, is countercyclical and large, 12.8 pp of annualized forward return separating the highest and lowest VIX quintiles and 25 to 30 pp over the year after a crisis-trough de-risk, and it is a return premium for bearing crisis risk rather than a Sharpe premium. Fourth, the part of TPA worth preserving is dynamic risk management, not concentration, and it needs no security selection; the finding reconciles with Moreira and Muir (2017, 2019), who cut on the rising-vol leg where the leverage effect dominates, the two policies being sequential complements.

What changes Monday morning is the reporting. Boards should request σ(TE) and the cyclicality of TE alongside return, volatility, and Sharpe, and read them against a stated drawdown budget. The level of $\tau_{max}$ and the cyclicality of TE are separate questions: a fund can have a wide budget and still fail to use it dynamically. Dynamic alpha is harvestable through a systematic regime signal without security selection, and the price of retreating to benchmark when stress peaks is the 12.8 pp omega premium at one month and the 25 to 30 pp forgone over the year after capitulating at the GFC and COVID troughs. Open questions remain: cross-country tests of convergence using funds with documented governance, contract-theoretic design of CIO incentives for a dynamic mandate, and the dynamic-alpha cost of illiquid private commitments. The community has spent two decades debating whether to adopt TPA when the productive question is how much drawdown the fund can bear and how dynamically the CIO is permitted to spend that budget. SAA and TPA are two steps toward answering it; the next is to budget the drawdown directly and treat tracking error, however dynamic, as the proxy it has always been.

## Appendix A: Multi-Asset Generalization

The exposition in Section 4 uses a scalar active tilt $\theta$ into a single risky asset. This appendix re-derives the central results, Hypotheses 1, 2, 3, and 5, in vector-matrix form, with a vector of active weights $\mathbf{w}$ and a full active-return covariance matrix $\mathbf{\Sigma}$. The generalization confirms that the scalar framework in the body is faithful to the multi-asset setting and does not depend on the single-bet abstraction.

A note on what these derivations are. They generalize the mean-variance benchmark of Section 4, the Treynor-Black optimum and its information ratio, to n assets. As Section 4.3 explains, that benchmark is a reference, not the policy the paper runs. On our data the information ratio does not rise in stress, so the mean-variance optimum would in fact lower tracking error there; the operative policy sizes tracking error from the board's drawdown budget, raising it in and after stress to hold the expected-return premium. The derivations below show the classical mean-variance results carry intact from one asset to n. The reason the policy departs from this optimum, onto the drawdown budget, is in the body, and the one place where the choice of ruler changes a sign, the shadow price, is flagged in Section A.5.

### A.1 Setup

Let the benchmark be a fixed policy portfolio with weights $\mathbf{b} \in \mathbb{R}^n$ across n risky assets and a possible riskless position. The portfolio held by the CIO is $\mathbf{b} + \mathbf{w}$, where $\mathbf{w} \in \mathbb{R}^n$ is the vector of active weights. Active weights sum to zero (cash-neutral active position): $\mathbf{1}'\mathbf{w} = \mathbf{0}$.

Let $\boldsymbol{\alpha}(t) \in \mathbb{R}^n$ denote the vector of expected active returns at time t (one per asset, in excess of the benchmark) and $\mathbf{\Sigma}(t) \in \mathbb{R}^{n\times n}$ denote the active-return covariance matrix, where $\mathbf{\Sigma}(t)$ is symmetric positive-definite. Both $\boldsymbol{\alpha}$ and $\mathbf{\Sigma}$ are functions of the regime, with values $\boldsymbol{\alpha}(H)$, $\mathbf{\Sigma}(H)$ in the high-risk state and $\boldsymbol{\alpha}(L)$, $\mathbf{\Sigma}(L)$ in the low-risk state.

Tracking error at time t is the standard deviation of active returns:

$$\mathrm{TE}(t) = \sqrt{(\mathbf{w}'\boldsymbol{\Sigma}(t)\mathbf{w})}$$

The portfolio's compound active return in period t, ignoring third-order terms, is:

$$r(t) = \mathbf{w}'\boldsymbol{\alpha}(t) - \tfrac{1}{2}\,\mathbf{w}'\boldsymbol{\Sigma}(t)\mathbf{w}$$

The first term is the linear reward; the second is the variance drain that imposes a quadratic penalty on the size of the active position. As in the scalar case, this penalty is what makes increasing tracking error costly even when the linear reward is positive.

**A.2 Unconstrained Optimum (Hypothesis 1, vector form)**

The CIO maximizes the per-period compound active return over **w**:

$$\mathbf{w}'\boldsymbol{\alpha}(t) - \tfrac{1}{2}\,\mathbf{w}'\boldsymbol{\Sigma}(t)\mathbf{w}$$

subject to $\mathbf{1}'\mathbf{w} = \mathbf{0}$ (cash-neutral). The first-order condition gives the optimal unconstrained active weight:

$$\mathbf{w}{*}(t) = \boldsymbol{\Sigma}(t)^{-1}\boldsymbol{\alpha}(t) - \lambda_b\,\boldsymbol{\Sigma}(t)^{-1}\mathbf{1}$$

where $\lambda_b$ is the Lagrange multiplier on the cash-neutrality constraint. When the riskless position absorbs any imbalance (the standard case), $\lambda_b = 0$ and the result simplifies to $\mathbf{w}{*}(t) = \boldsymbol{\Sigma}(t)^{-1}\boldsymbol{\alpha}(t)$. This is the textbook multi-asset Markowitz (1952) / Treynor-Black (1973) active solution; it reduces to the scalar $\theta{*} = \alpha/\sigma^2$ when $n = 1$.

The optimal unconstrained tracking error is:

$$\mathrm{TE}^{*}(t) = \sqrt{(\mathbf{w}{*}'\boldsymbol{\Sigma}(t)\mathbf{w}{*})} = \sqrt{(\boldsymbol{\alpha}(t)'\boldsymbol{\Sigma}(t)^{-1}\boldsymbol{\alpha}(t))}$$

The quantity $\sqrt{(\boldsymbol{\alpha}'\boldsymbol{\Sigma}^{-1}\boldsymbol{\alpha})}$ is the multi-asset information ratio, the natural generalization of the scalar $\alpha/\sigma$. In this mean-variance benchmark the optimal tracking error scales with that ratio, so it rises across regimes only if the ratio does. On our data the ratio does not rise in stress (Section 4.3), so the benchmark optimum would lower tracking error there; the body's policy instead sizes tracking error from the drawdown budget, raising it in and after stress to hold the expected-return premium that the lower information ratio still leaves on the table.

**A.3 Constrained Optimum (Hypothesis 2, vector form)**

The TE constraint $\tau_{max}$ in the multi-asset setting is:

$$\mathbf{w}'\mathbf{\Sigma}(t)\mathbf{w} \leq \tau_{max}^2$$

Forming the Lagrangian:

$$\mathcal{L}(\mathbf{w}, \lambda) = \mathbf{w}'\boldsymbol{\alpha}(t) - ½\, \mathbf{w}'\mathbf{\Sigma}(t)\mathbf{w} - ½\, \lambda\, (\mathbf{w}'\mathbf{\Sigma}(t)\mathbf{w} - \tau_{max}^2)$$

The first-order condition gives:

$$\mathbf{w}*(t) = \mathbf{\Sigma}(t)^{-1}\boldsymbol{\alpha}(t) \,/\, (1 + \lambda)$$

with complementary slackness $\lambda \geq 0$, $\lambda(\mathbf{w}*'\mathbf{\Sigma}\mathbf{w}* - \tau_{max}^2) = 0$. When the constraint binds, $\lambda > 0$ and we solve:

$$\tau_{max}^2 = (1 + \lambda)^{-2}\, \boldsymbol{\alpha}(t)'\mathbf{\Sigma}(t)^{-1}\boldsymbol{\alpha}(t)$$

which gives:

$$1 + \lambda = \sqrt{(\boldsymbol{\alpha}(t)'\mathbf{\Sigma}(t)^{-1}\boldsymbol{\alpha}(t))} \,/\, \tau_{max} = \text{TE}^*(t) \,/\, \tau_{max}$$

So the constrained active weight is just the unconstrained weight scaled down by the ratio of unconstrained-to-cap tracking error:

$$\mathbf{w}*(t) = (\tau_{max} \,/\, \text{TE}^*(t)) \times \mathbf{\Sigma}(t)^{-1}\boldsymbol{\alpha}(t)$$

when $\text{TE}*(t) > \tau_{max}$, and $\mathbf{w}*(t) = \mathbf{\Sigma}(t)^{-1}\boldsymbol{\alpha}(t)$ otherwise. The realized tracking error is therefore:

$$TE(t) = \min(TE^*(t), \tau_{max})$$

This is the same form as the scalar case in the body of the paper. The volatility of realized TE, σ(TE), is large when $\tau_{max}$ is loose (rarely binds) and small when $\tau_{max}$ is tight (binds in most states), reproducing Hypothesis 2 in the multi-asset case.

**A.4 Compound Return Advantage (Hypothesis 3, vector form)**

The compound active return in any single period at the unconstrained optimum, obtained by substituting $\mathbf{w}*$ into r(t), is:

$$r^*(t) = \tfrac{1}{2}\, \boldsymbol{\alpha}(t)'\boldsymbol{\Sigma}(t)^{-1}\boldsymbol{\alpha}(t) = \tfrac{1}{2}\, IR(t)^2$$

where $IR(t) = \sqrt{(\boldsymbol{\alpha}(t)'\boldsymbol{\Sigma}(t)^{-1}\boldsymbol{\alpha}(t))}$ is the multi-asset information ratio in regime t.

The dynamic portfolio that adapts to each regime earns expected compound advantage:

$$E[r^*] = \tfrac{1}{2}\, \{p\ IR(H)^2 + (1 - p)\ IR(L)^2\}$$

The static portfolio that uses a fixed information ratio across regimes earns ½ × (blended IR)$^2$. The dynamic advantage is:

$$\Delta = \tfrac{1}{2}\, \{E[IR^2] - (E[IR])^2\} > 0$$

The result holds whenever IR genuinely varies across states. The Jensen's inequality channel operates in the vector case for exactly the same reason as the scalar: squaring is convex.

The intuition extends: a CIO with no security-selection skill at all (no superior $\boldsymbol{\alpha}$ estimates) can still earn this mean-variance timing advantage simply by varying $\mathbf{w}*$ with $\boldsymbol{\alpha}(t)$ and $\boldsymbol{\Sigma}(t)$ as they shift across regimes. It scales with the *dispersion* of $IR^2$ across states, regardless of asset-count dimension.

This ½ Var(IR) term is the mean-variance benchmark's timing term, and it rewards adapting to the information ratio, which on our data is higher in calm than in stress. The policy the paper runs does not chase it. Its compound advantage comes from the drawdown-budget channel of Section 4.3, carrying more of the equity premium while a floor holds the worst drawdown to the board's budget, which is why the realized policy raises rather than lowers tracking error in stress.

**A.5 The Shadow Price Omega (Hypothesis 5, vector form)**

The multi-asset constraint is $\sqrt{(\mathbf{w}'\boldsymbol{\Sigma}(t)\mathbf{w})} \leq \tau_{max}$, written as a linear constraint on tracking error rather than as a quadratic constraint on squared TE. The Lagrangian is:

$$\mathcal{L}(\mathbf{w}, \lambda) = \mathbf{w}'\boldsymbol{\alpha}(t) - \tfrac{1}{2}\, \mathbf{w}'\boldsymbol{\Sigma}(t)\mathbf{w} - \lambda \times (\sqrt{(\mathbf{w}'\boldsymbol{\Sigma}(t)\mathbf{w})} - \tau_{max})$$

with first-order condition $\boldsymbol{\alpha}(t) = \boldsymbol{\Sigma}(t)\mathbf{w} \times (1 + \lambda / \tau_{max})$. At the binding constraint $\sqrt{(\mathbf{w}'\boldsymbol{\Sigma}(t)\mathbf{w})} = \tau_{max}$, the active weight is $\mathbf{w}* = (\tau_{max} / (\tau_{max} + \lambda))\, \boldsymbol{\Sigma}(t)^{-1}\boldsymbol{\alpha}(t)$. Substituting back into the binding constraint gives:

$$\tau_{max} + \lambda = \sqrt{(\boldsymbol{\alpha}(t)'\boldsymbol{\Sigma}(t)^{-1}\boldsymbol{\alpha}(t))} = \mathrm{TE}^*(t)$$

so the shadow price in tracking-error units is:

$$\lambda(t) = \mathrm{TE}^*(t) - \tau_{max} = \sqrt{(\boldsymbol{\alpha}(t)'\boldsymbol{\Sigma}(t)^{-1}\boldsymbol{\alpha}(t))} - \tau_{max}$$

For $n = 1$, $\boldsymbol{\Sigma} = \sigma^2$, $\boldsymbol{\alpha}'\boldsymbol{\Sigma}^{-1}\boldsymbol{\alpha} = \alpha^2/\sigma^2$, and $\lambda(t)$ collapses to $\alpha(t)/\sigma(t) - \tau_{max}$. This is exactly the information-ratio-units multiplier reported in the scalar derivation of the body (Section 4.3, line preceding the definition of $\Omega$). The body then converts to expected-return units by multiplying by $\sigma(t)$:

$$\Omega_{body}(t) = \sigma(t) \times \lambda_{scalar}(t) = \alpha(t) - \tau_{max}\, \sigma(t)$$

The multi-asset analogue, in expected-return-on-the-binding-portfolio units, is:

$$\Omega(t) = \tau_{max} \times \lambda(t) = \tau_{max} \times (TE^*(t) - \tau_{max})$$

which is the expected one-period return the constrained CIO forgoes by holding TE at $\tau_{max}$ rather than at $TE*(t)$. For n = 1, this collapses to $\tau_{max} \times (\alpha/\sigma - \tau_{max}) = \tau_{max}\alpha/\sigma - \tau_{max}^2$, a related but distinct quantity from the body's $\Omega = \alpha - \tau_{max}\sigma$. The two are different functions, not rescalings of each other, and under the regime data they part ways. The body's $\alpha - \tau_{max}\sigma$ is the marginal alpha at the constraint and rises in stress with the expected return $\alpha$; $\tau_{max} \times \lambda$ tracks the information ratio $\alpha/\sigma$, which on our data does not rise in stress. We report and rely on the body's marginal-alpha form, which is the one that spikes when forward returns are richest and the constraint bites hardest.

The countercyclical behavior holds in the marginal-alpha form the body uses, and not in the information-ratio form. Because the expected active return $\alpha$ is large in the high-risk state, the body's omega $\Omega = \alpha - \tau_{max}\sigma$ spikes there, exactly as in Section 4.3, even though the information ratio $\alpha/\sigma$ does not rise. The constrained investor forgoes the largest expected return precisely when forward returns are highest. This is the multi-asset analogue of the Brunnermeier-Pedersen (2009) funding-liquidity channel and the Alankar-Blaustein-Scholes (2013) characterization of the joint TE-liquidity shadow price.

### A.6 Mapping to the Scalar Exposition of Section 4

Setting n = 1, the benchmark reduces to the scalar forms used in the body, with two rows where the body's operative policy departs from the mean-variance benchmark:

| **Body** | **Appendix** |
|---|---|
| $\theta^* = \alpha/\sigma^2$ | $\mathbf{w}* = \mathbf{\Sigma}^{-1}\boldsymbol{\alpha}$ |
| $TE^* = \alpha/\sigma$ | $TE^* = \sqrt{(\boldsymbol{\alpha}'\mathbf{\Sigma}^{-1}\boldsymbol{\alpha})}$ |
| $TE = \min(TE_{target}, \tau_{max})$ | $TE = \min(TE^*, \tau_{max})$ |
| compound advantage via the drawdown budget (§4.3) | ½ $\boldsymbol{\alpha}'\boldsymbol{\Sigma}^{-1}\boldsymbol{\alpha}$ (benchmark) |
| $\lambda_{scalar} = \alpha/\sigma - \tau_{max}$ (IR units) | $\lambda = TE^* - \tau_{max}$ (TE units) |
| $\Omega = \alpha - \tau_{max}\,\sigma = \sigma \times \lambda_{scalar}$ | $\Omega = \tau_{max} \times \lambda$ (forgone active return) |

The scalar mean-variance exposition in the body is a faithful one-asset projection of this benchmark. Two rows differ because the body's operative policy departs from the benchmark: it sets realized tracking error from the policy target $TE_{target}$ rather than from the unconstrained optimum, and it draws its compound advantage from the drawdown budget (Section 4.3) rather than from the dispersion of the information ratio. The shadow price differs in form as well, as Section A.5 details, and the body relies on the marginal-alpha Ω.

The multi-asset extension does add one substantive point. In the vector case, $\boldsymbol{\Sigma}(t)$ can shift even when individual asset volatilities are unchanged, through correlation regime changes (Exhibit 1 and Exhibit 2 of the body document exactly this). The information ratio depends on $\boldsymbol{\Sigma}^{-1}$, so a correlation spike that leaves marginal volatilities unchanged can still dramatically alter IR(t) and therefore the optimal tracking error. This is the formal counterpart to the body's argument that "stocks become a stock market" during stress: the rise in pairwise correlations changes $\boldsymbol{\Sigma}$ and the active-portfolio problem in a way the scalar framework cannot represent. The vector generalization is therefore not just rhetorically reassuring; it is the natural setting in which the regime-switching opportunity set is fully expressed.

# Exhibits

## Exhibit 1: Rolling 63-Day Average Pairwise Sector Correlation (2000–2026)

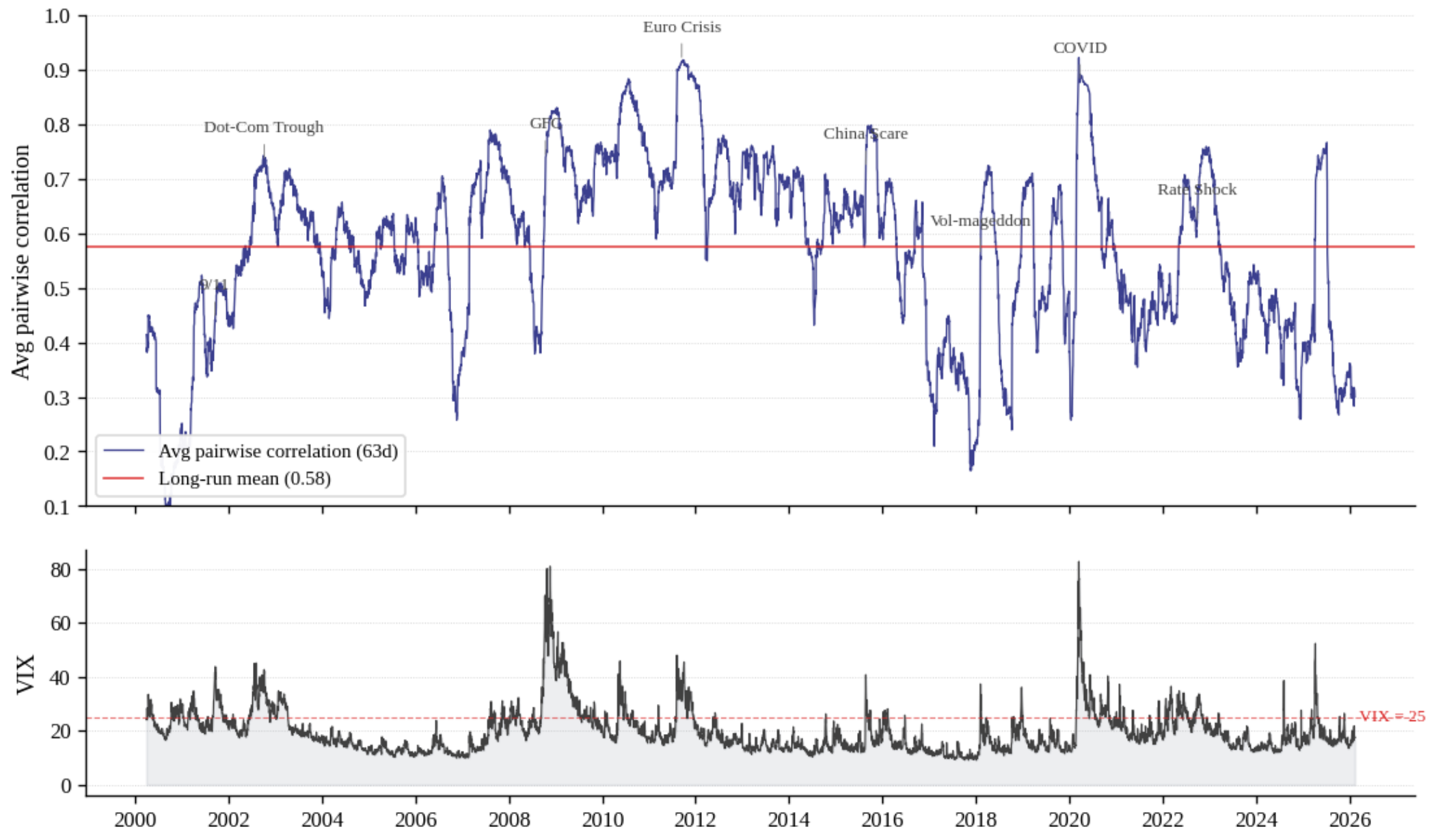

## Exhibit 2: Rolling Stock-Bond Correlation (2003–2026)

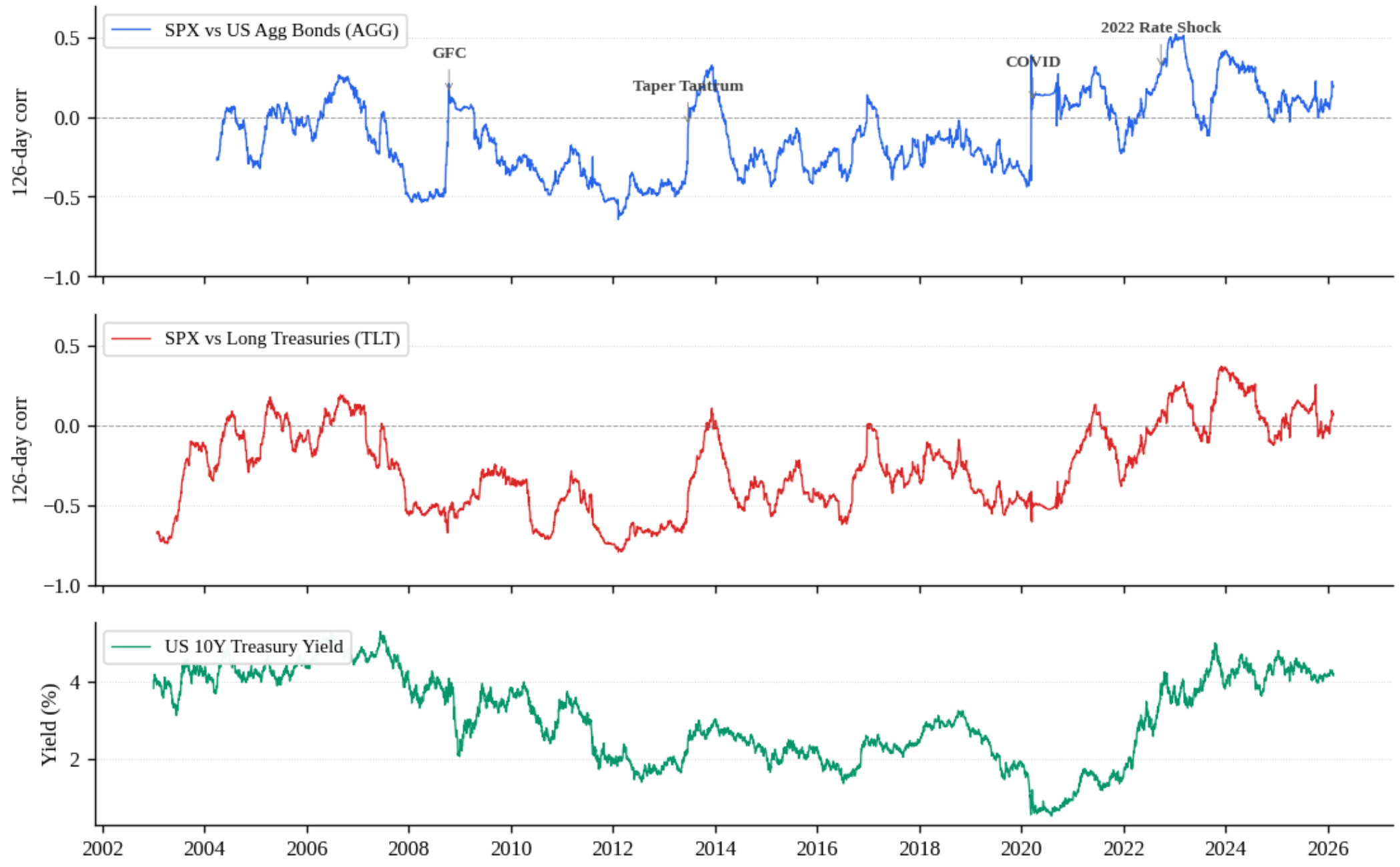

**Exhibit 3: Static Versus Dynamic Tracking Error, Cumulative Active Return and Realized TE Paths**

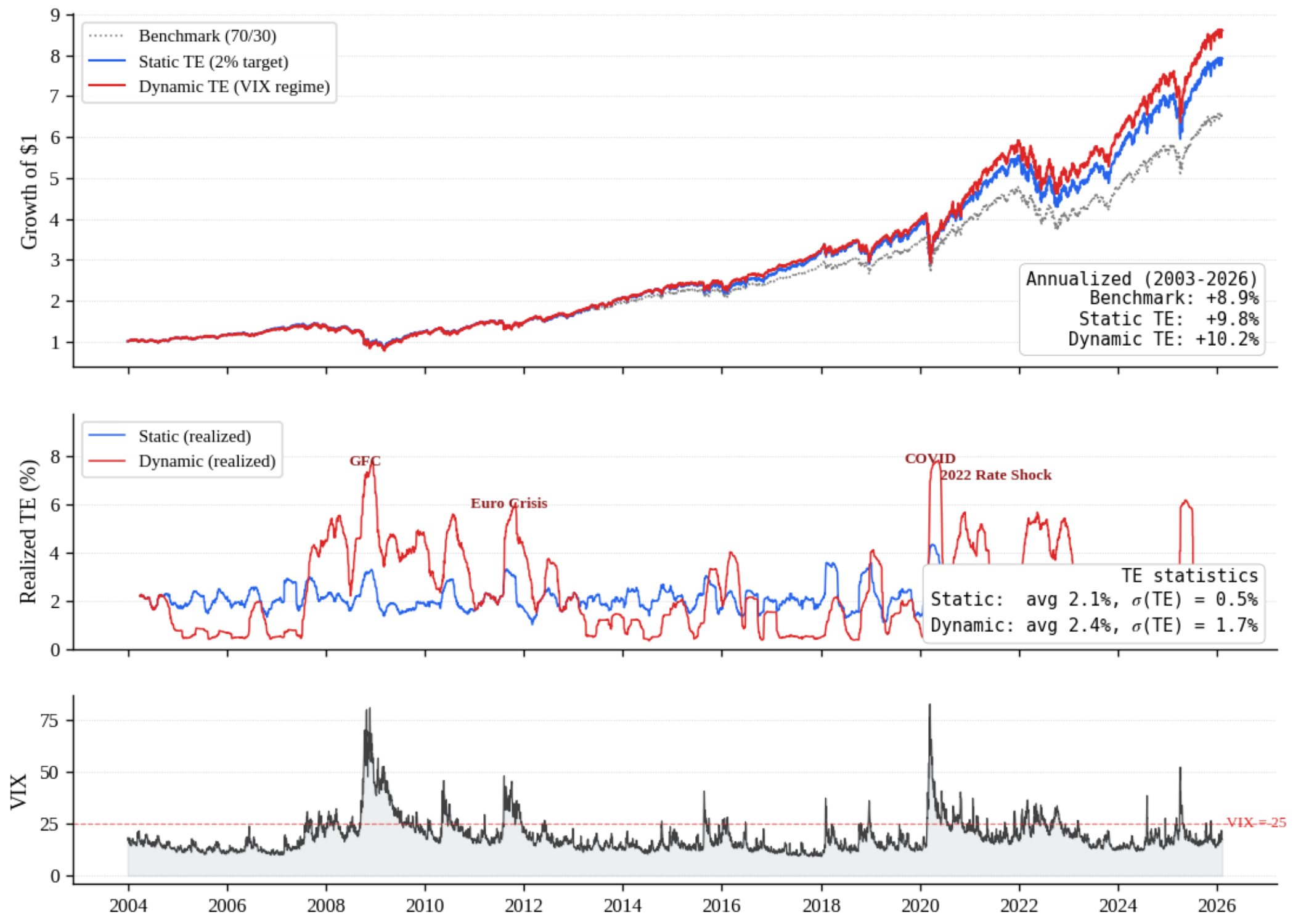

**Exhibit 4, Panel A: The Omega Premium, VIX Quintile Forward Returns (1990–2026)**

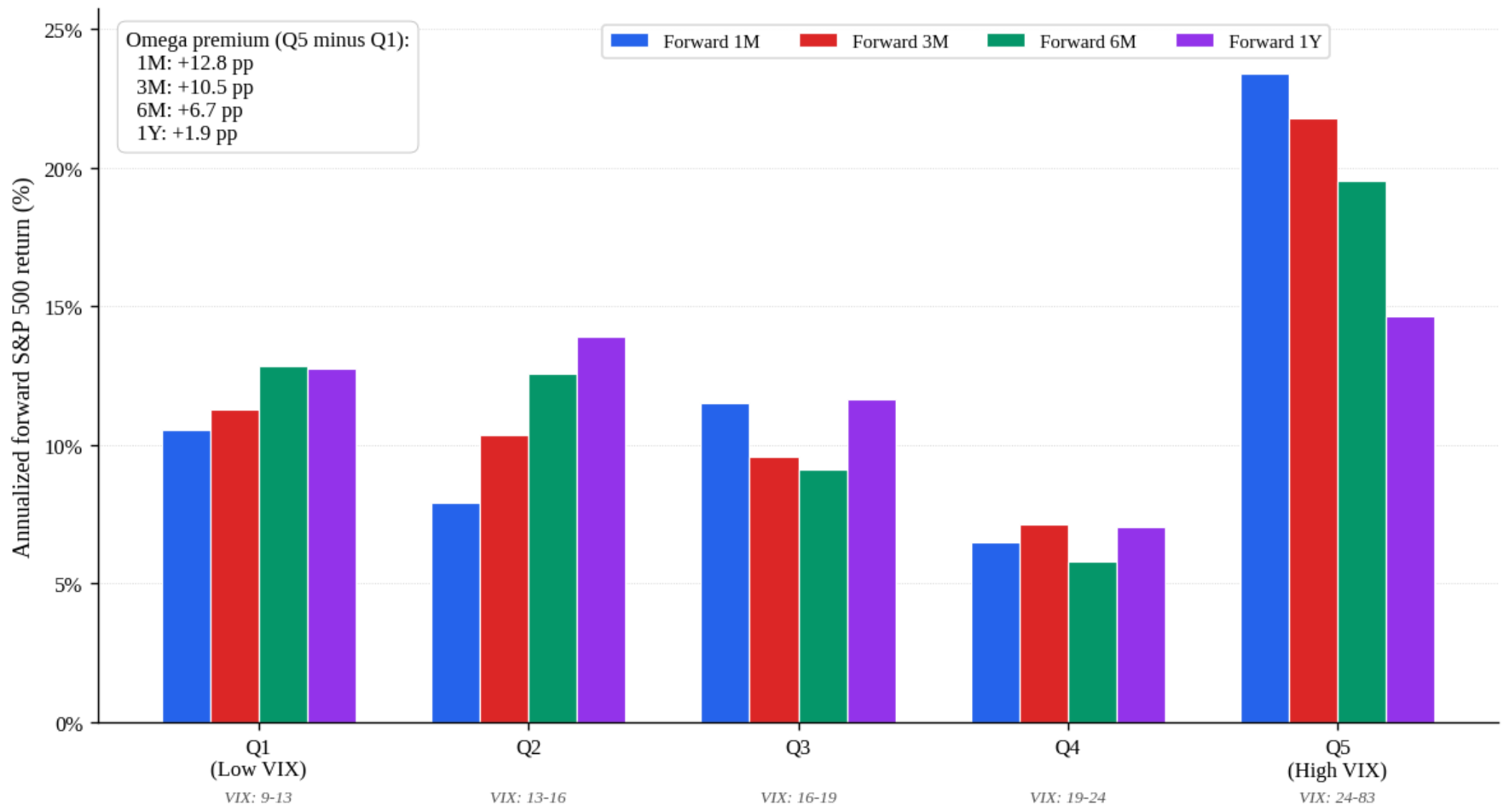

## Exhibit 4, Panel B: The Omega Premium by Horizon

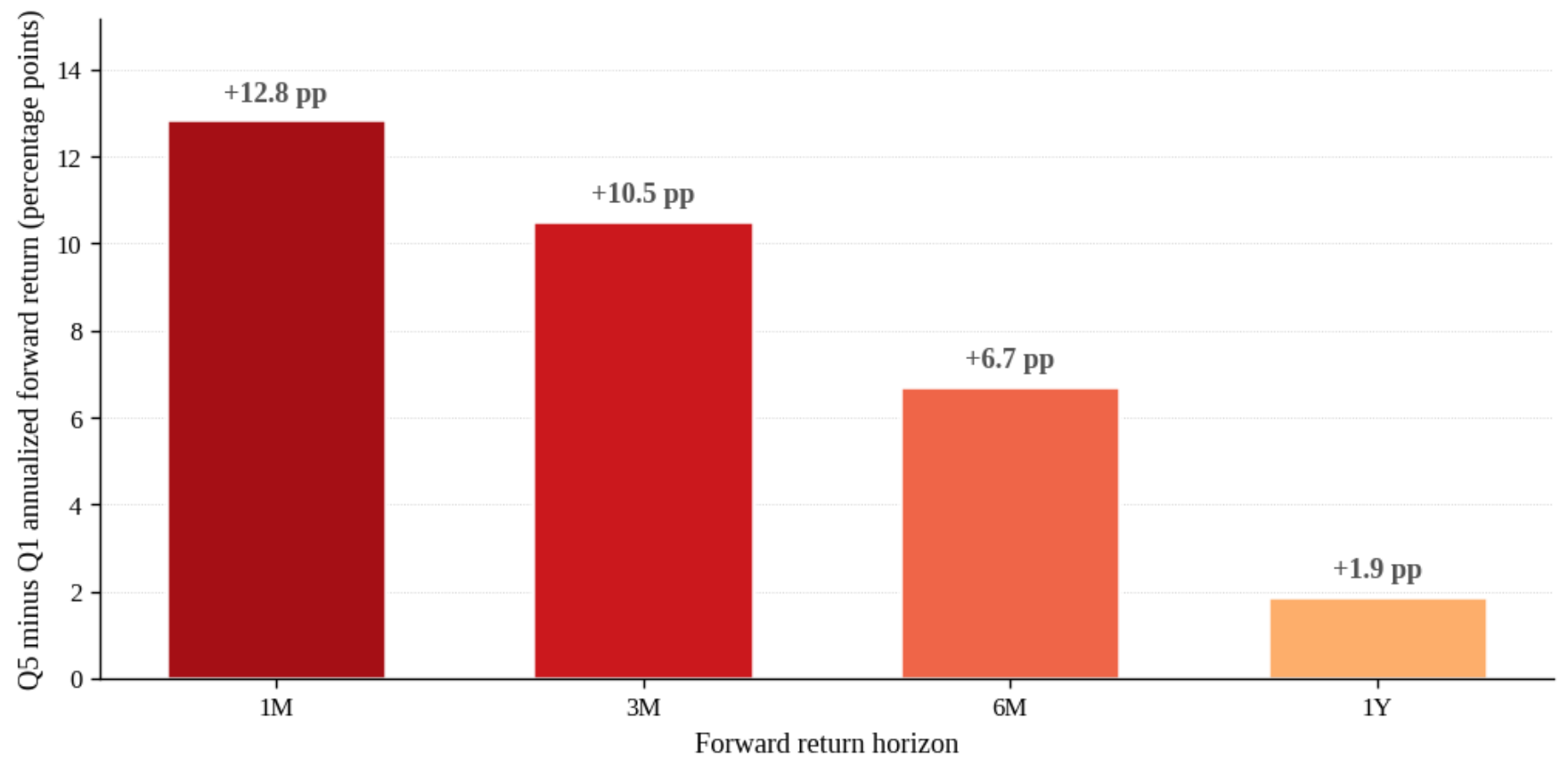

*Spread between top-VIX quintile (Q5) and bottom-VIX quintile (Q1) annualized forward SPX returns. The premium decays with horizon, consistent with short-horizon liquidity provision rather than a persistent factor.*

## Exhibit 4, Panel C: The Omega Premium in Return and in Sharpe, by VIX Quintile (Forward One Month)

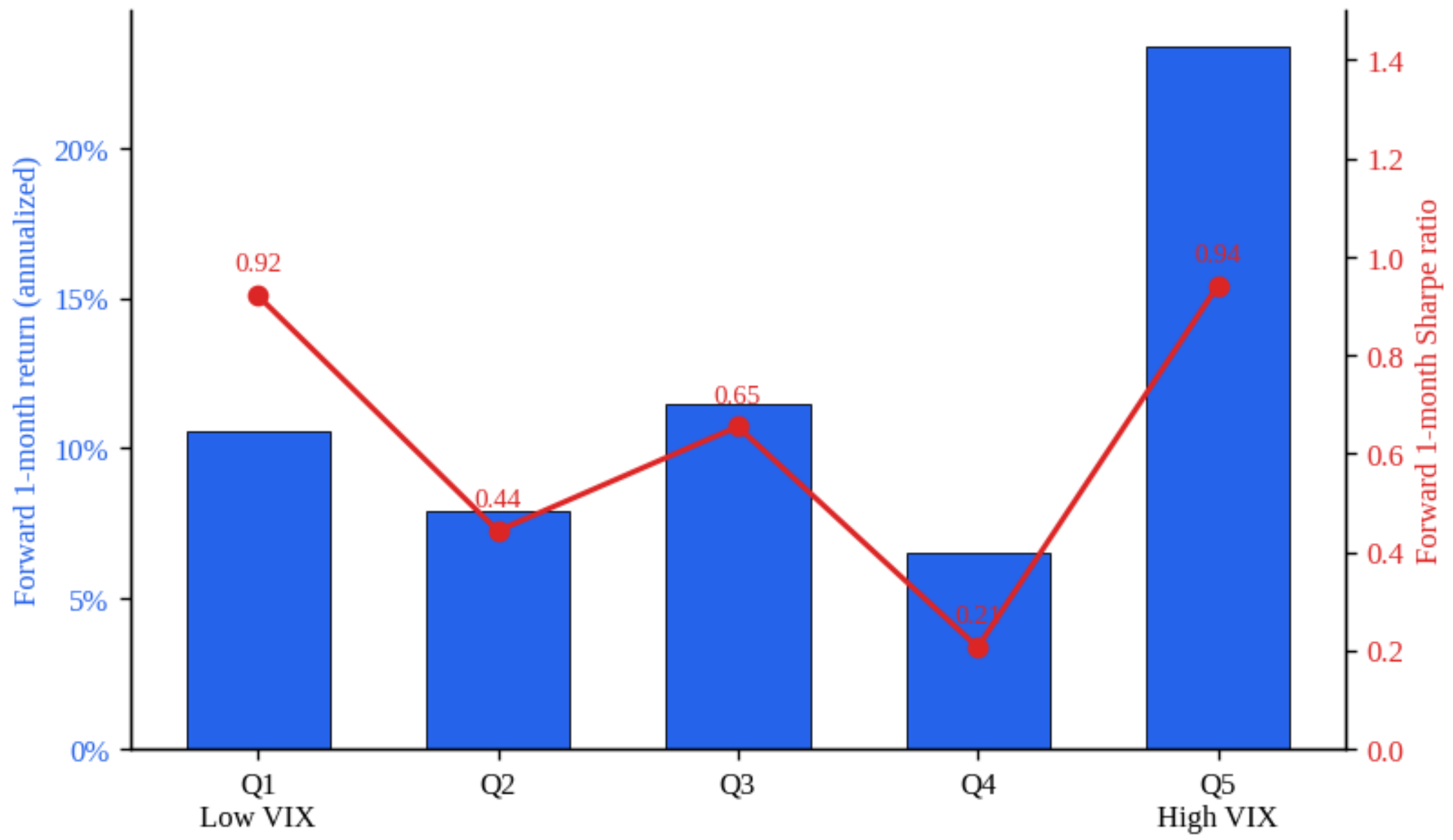

## Exhibit 5, Panel A: Portfolio Drawdown, VIX, and Average Sector Correlation (2003–2026)

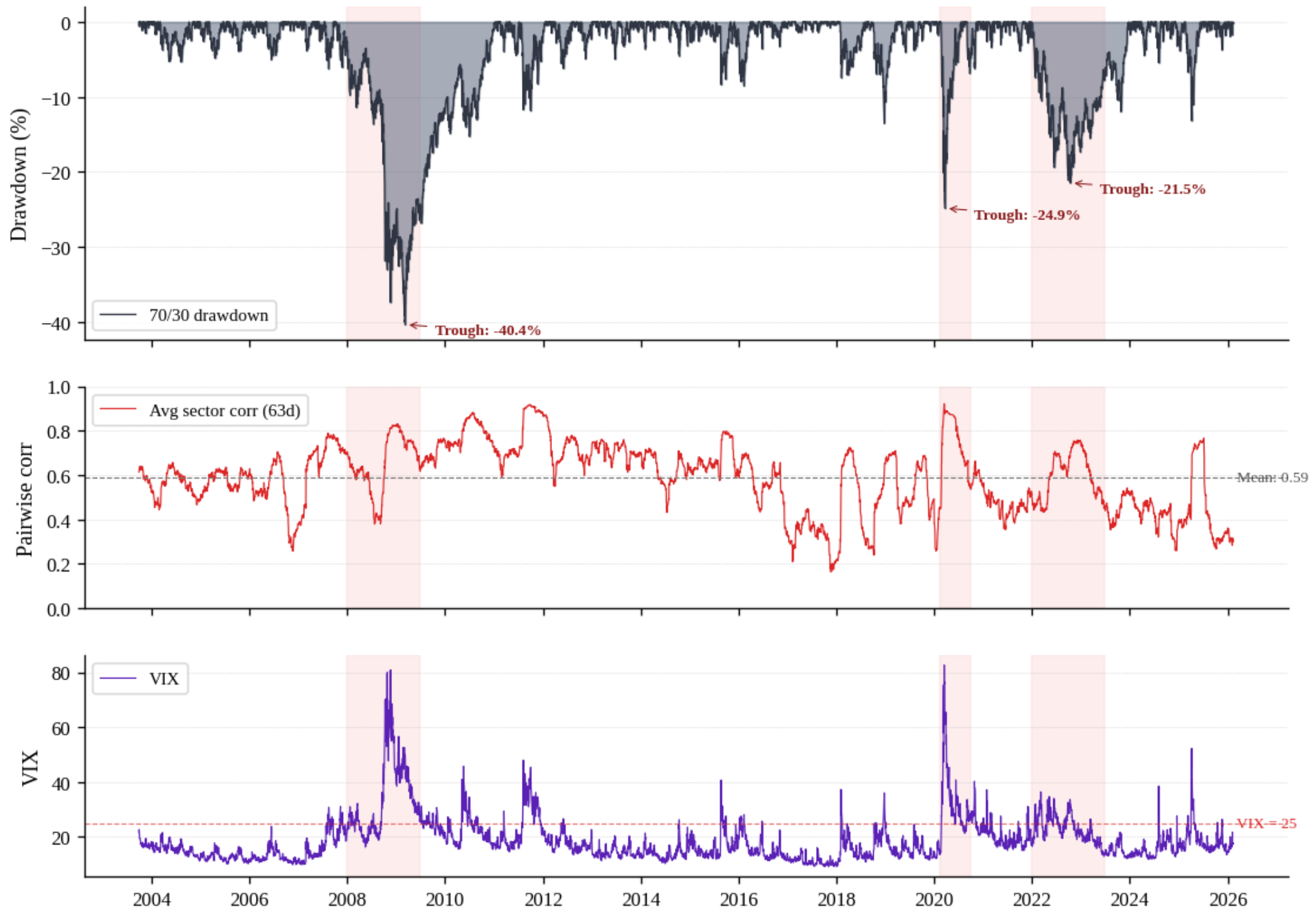

## Exhibit 5, Panel B: The Cost of De-Risking at the Trough

| Crisis | Max DD | VIX | Horizon | Stay 70/30 | De-risk 30/70 | Regret |
|---|---|---|---|---|---|---|
| 2008 GFC | −40.4% | 49.7 | 3M | 27.1% | 11.5% | 15.6 pp |
| 2008 GFC | −40.4% | 49.7 | 6M | 36.9% | 18.4% | 18.5 pp |
| 2008 GFC | −40.4% | 49.7 | 12M | 50.9% | 25.5% | 25.4 pp |
| 2020 COVID | −24.9% | 61.6 | 3M | 29.0% | 15.0% | 14.0 pp |
| 2020 COVID | −24.9% | 61.6 | 6M | 34.6% | 17.8% | 16.7 pp |
| 2020 COVID | −24.9% | 61.6 | 12M | 52.0% | 22.3% | 29.6 pp |
| 2022 Rate Shock | −21.4% | 33.6 | 3M | 10.1% | 7.7% | 2.4 pp |
| 2022 Rate | −21.4% | 33.6 | 6M | 13.5% | 9.0% | 4.4 pp |
| Shock | | | | | | |
| 2022 Rate Shock | −21.4% | 33.6 | 12M | 16.3% | 7.5% | 8.8 pp |

## Exhibit 5, Panel C: Forgone Return from Forced De-Risking at Crisis Trough

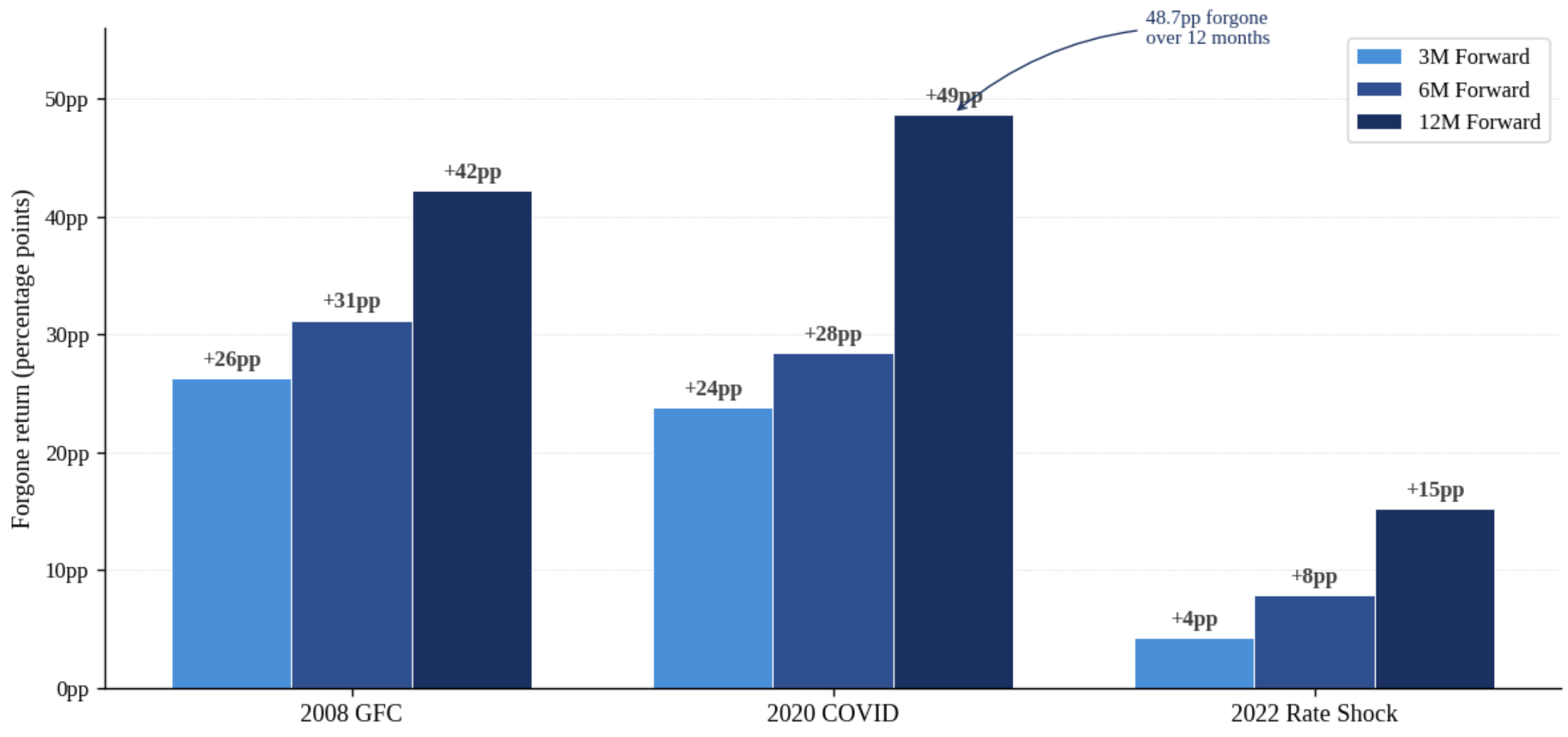

*Return differential: staying 70/30 versus moving to all bonds, measured from each crisis trough.*

# Exhibit 6, Panel A: TPA-SAA Convergence Under Constraints

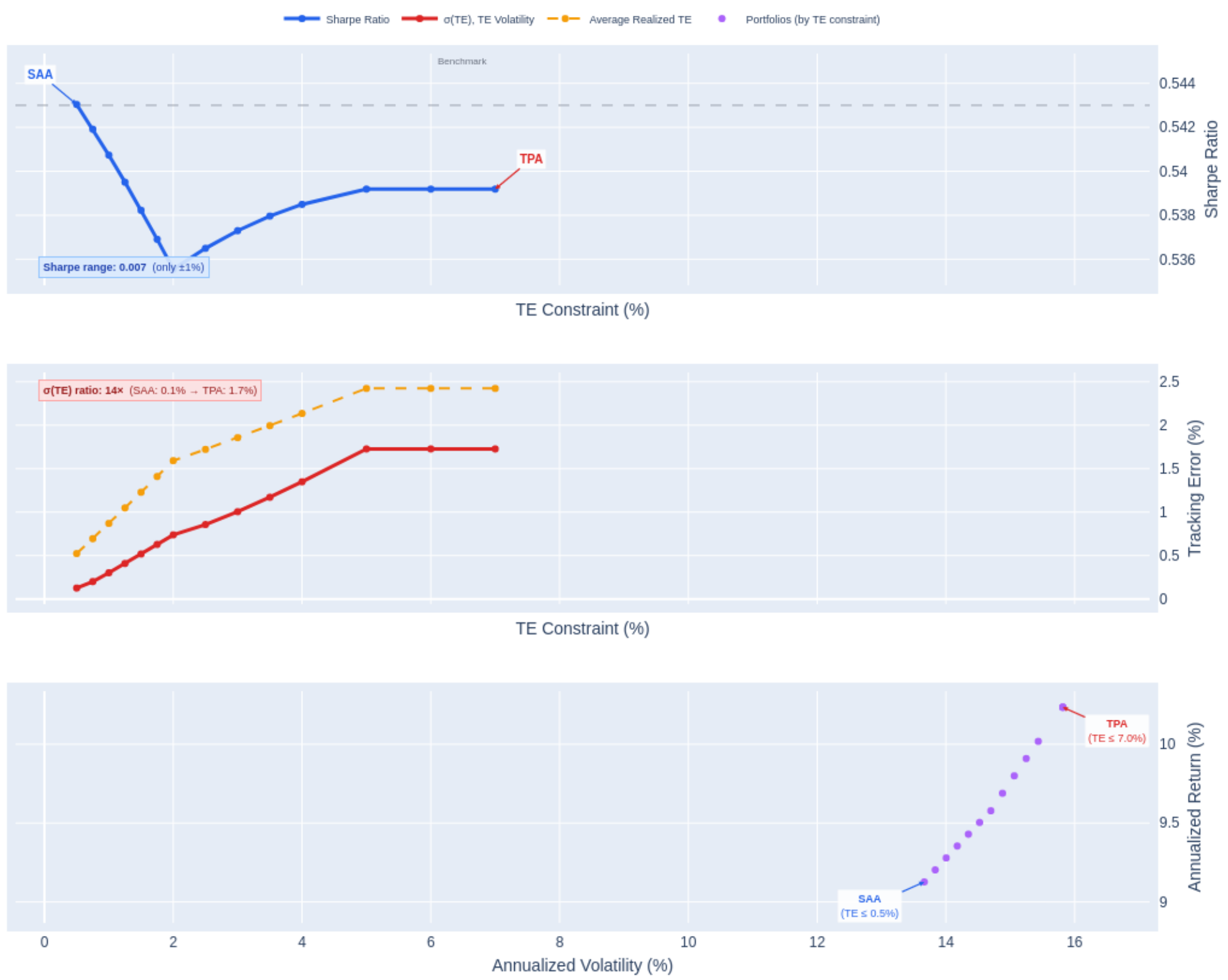

## Exhibit 6, Panel B: Realized Tracking Error Across Constraint Regimes

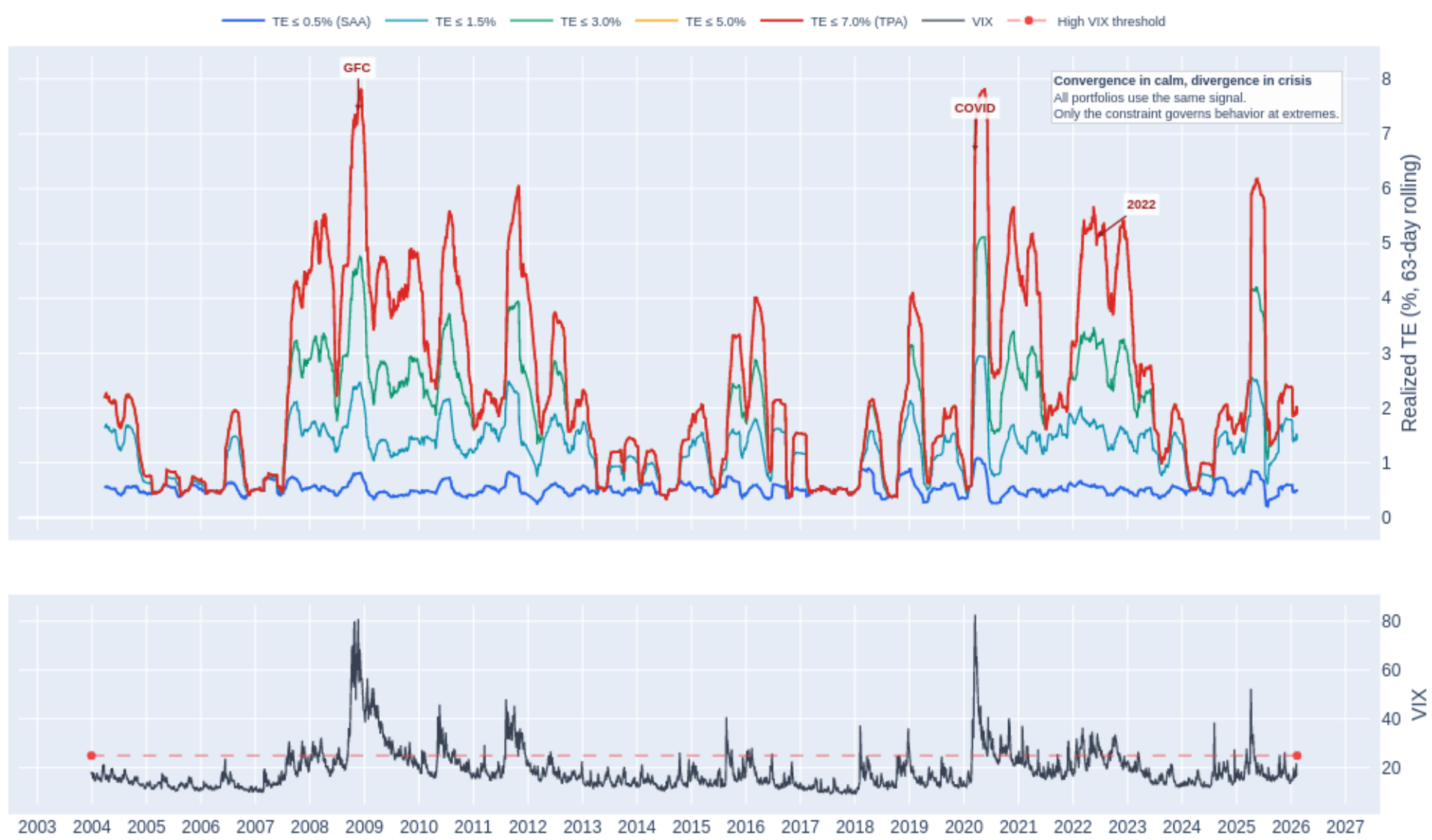

## Exhibit 7, Panel A: Cumulative Returns of Buy-and-Hold, Moreira-Muir, and TPA Overlay

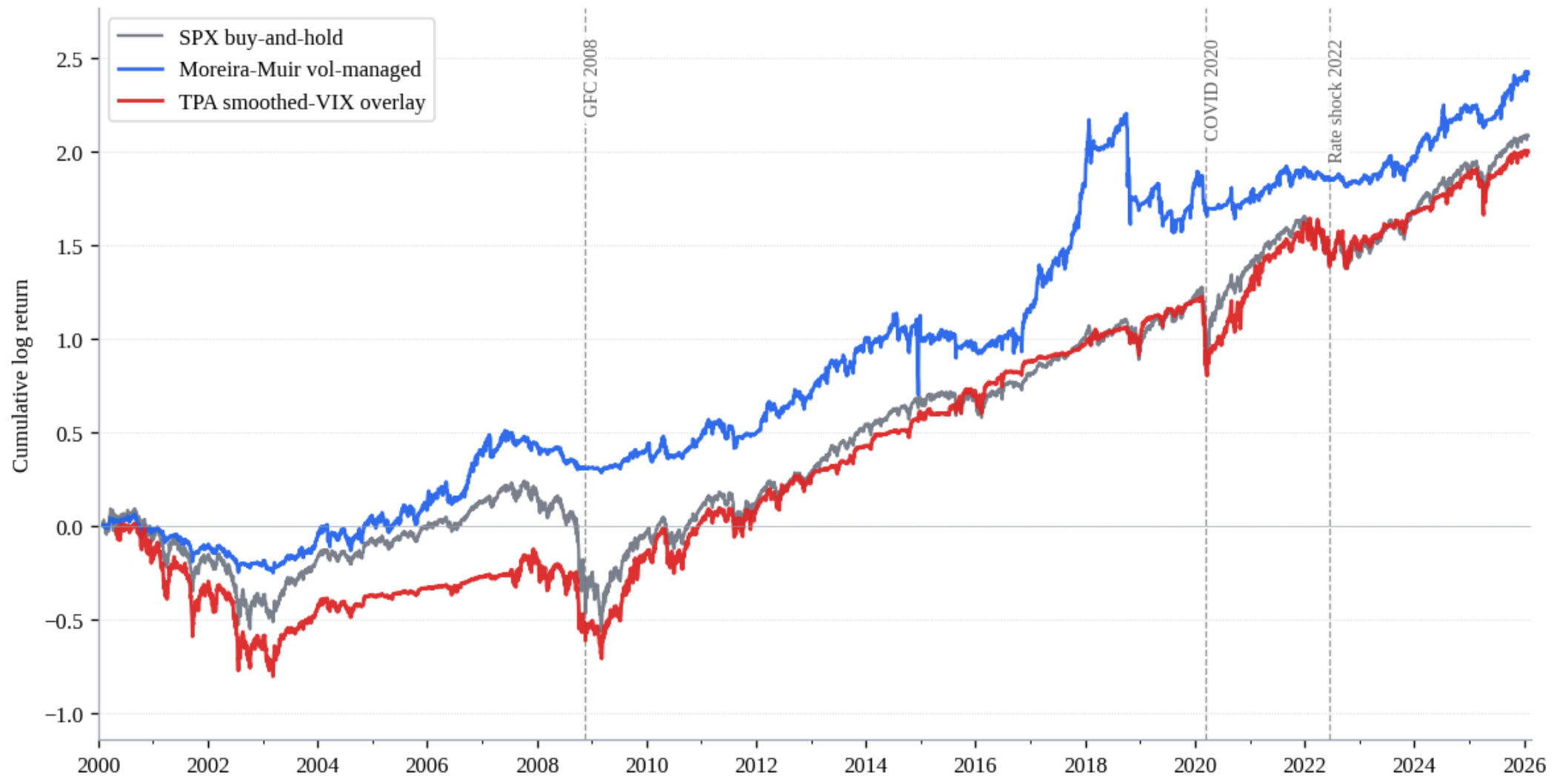

*Sharpe: BH 0.335 | MM 0.407 (full sample) | TPA-overlay 0.532 (2003-2026 overlap)*

## Exhibit 7, Panel B: Returns by Vol-Cycle Phase

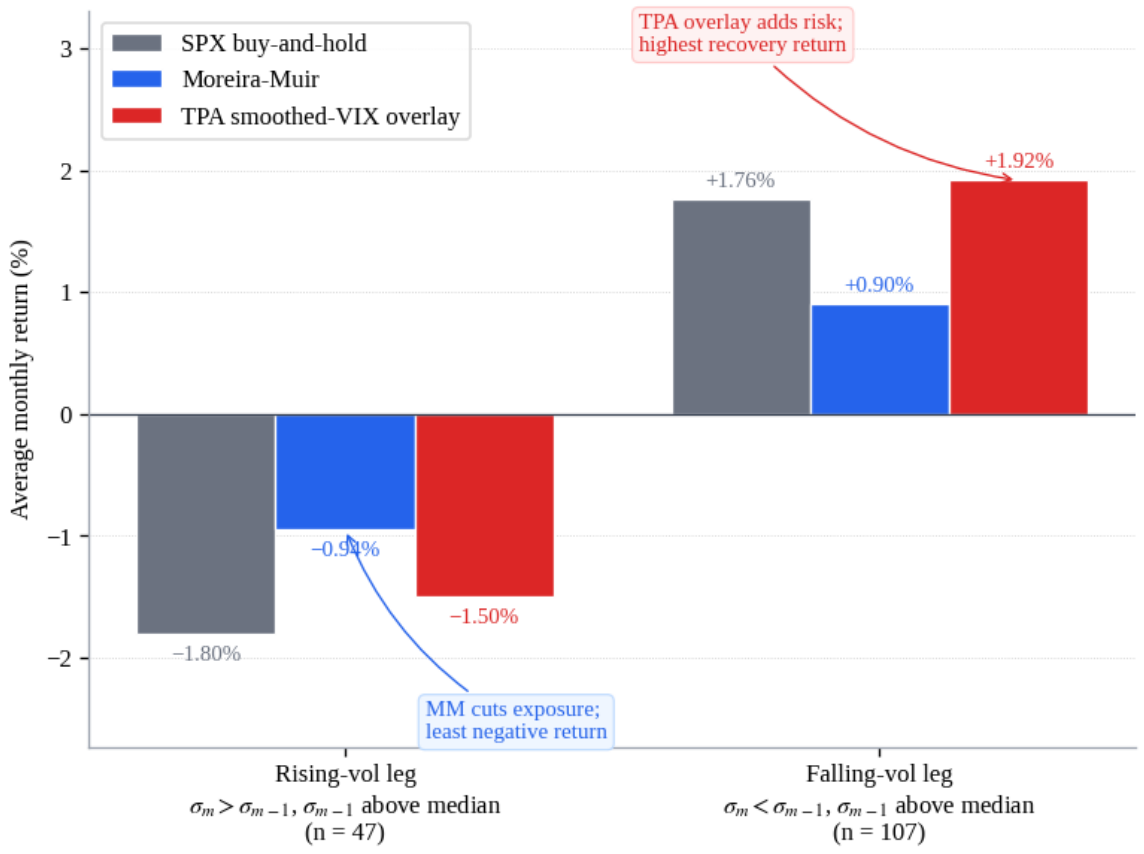

*Months with elevated lagged volatility, split by within-month vol surprise. MM mitigates the rising-vol leg; the TPA overlay lifts on the falling-vol leg. The two strategies occupy different halves of the same volatility cycle.*

## Exhibit 9, Panel A: Robustness Across Signals, Windows, and Thresholds ($\tau_{max}$ = 5%)

| | 10-day | 21-day | 42-day | 63-day |
|---|---|---|---|---|
| VIX | | | | |
| Sharpe range | 0.479–0.502 | 0.469–0.501 | 0.484–0.496 | 0.483–0.502 |
| σ(TE) range | 1.38%–1.59% | 1.24%–1.52% | 1.07%–1.48% | 0.99%–1.44% |
| Omega prem. | +15.6 pp | +15.9 pp | +10.6 pp | +10.2 pp |
| RV | | | | |
| Sharpe range | 0.461–0.482 | 0.460–0.475 | 0.463–0.474 | 0.467–0.489 |
| σ(TE) range | 1.45%–1.65% | 1.37%–1.69% | 1.21%–1.65% | 1.15%–1.56% |
| Omega prem. | +3.4 pp | +3.8 pp | +8.0 pp | +7.7 pp |
| BAA-10Y | | | | |
| Sharpe range | 0.488–0.494 | 0.489–0.490 | 0.474–0.484 | 0.475–0.495 |
| σ(TE) range | 1.16%–1.62% | 1.13%–1.60% | 1.08%–1.59% | 1.06%–1.50% |
| Omega prem. | +4.2 pp | +1.4 pp | −1.4 pp | −1.5 pp |

# Exhibit 9, Panel B: Omega Premium by Signal and Window

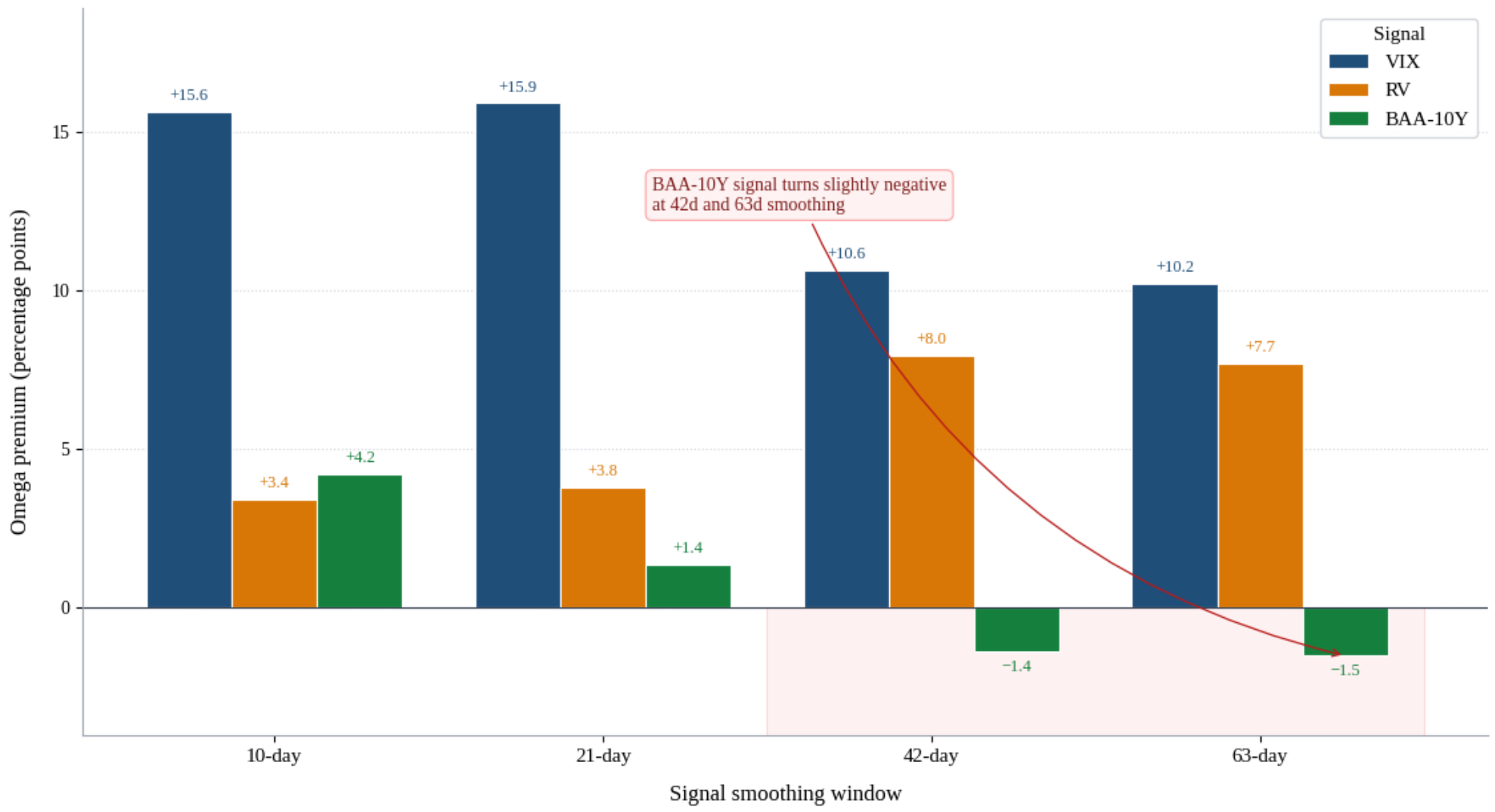

*Top-quintile minus bottom-quintile annualized forward 1-month SPX return. Aligned 2003-2026 sample. Omega depends only on (signal, window); threshold pairs do not affect this quintile-on-signal metric.*

# Exhibit 10: Net Sharpe Ratios After Transaction Costs

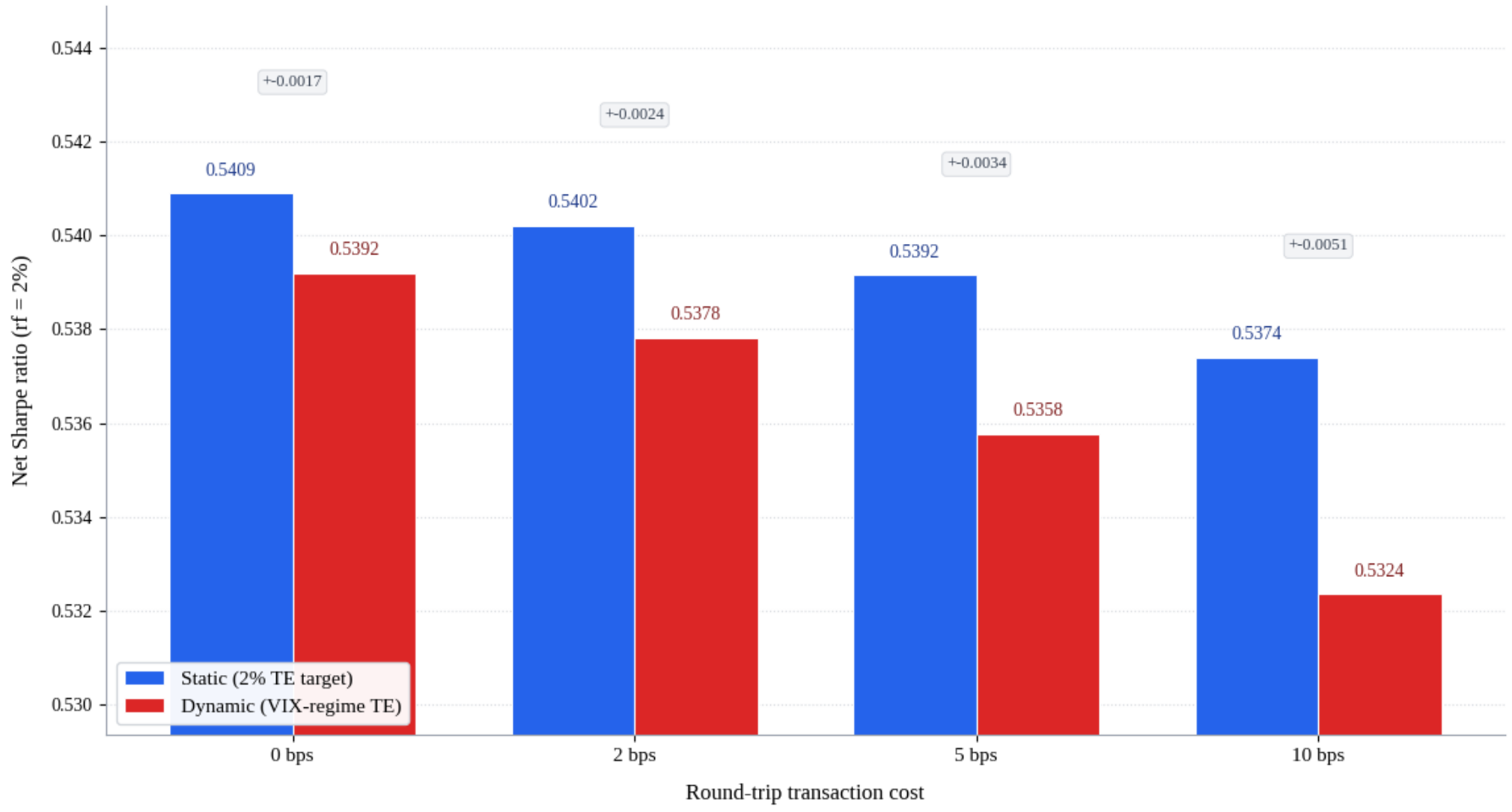

*Sharpe gap shrinks from -0.0017 (gross) to -0.0051 at 10 bps round-trip.*

**Exhibit 11: Statistical Tests of the Five Hypotheses**

| Hypothesis | Null | Test | Statistic | 95% CI | p-value | Result |
|---|---|---|---|---|---|---|
| H1 | ρ(VIX, TE) = 0 | Pearson | ρ = 0.822; t = 107.0 | [0.774, 0.871] | < 0.001 | Reject $H_0$; H1 supported |
| H2 | σ(TE) equal across $\tau_{max}$ | Bartlett; max/min ratio | $\chi^2$ = 28,233; ratio = 13.7× | [11.9×, 16.1×] | < 0.001 | Reject $H_0$; H2 supported |
| H3 | $E[r_{dyn}] = E[r_{static}]$ | Paired Newey-West (lag 21) | t = 1.61; ann. diff = 0.50 pp | [−0.11, +1.12] pp | 0.107 | Fail to reject $H_0$ at 5%; sign consistent with H3 |
| H4 | Sharpe equal across $\tau_{max}$ | Jobson-Korkie / Memmel | max \|z\| = 0.46; diff(0.5–7%) = 0.004 | [−0.036, +0.042] | min p = 0.645 | Fail to reject $H_0$; consistent with H4 (convergence) |
| H5 | Forward return monotone in VIX | Spearman; Q5–Q1 spread | ρ = 0.125; spread_1M = 12.8 pp | [−0.6, 25.8] pp | < 0.001 | Reject $H_0$; H5 supported |

## Exhibit 12, Panel A: Compound Return at an Equal One-Year Drawdown Budget (~38%)

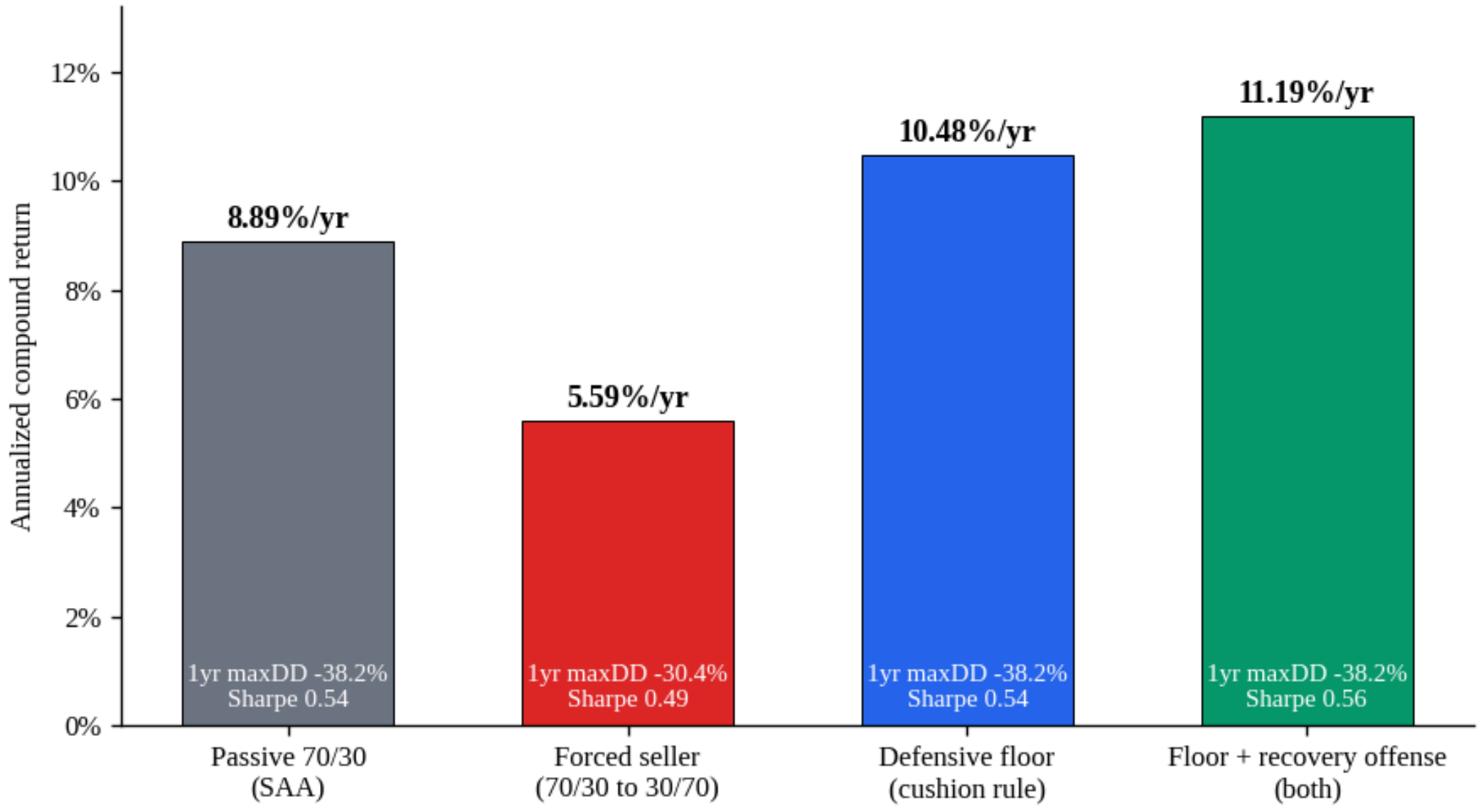

## Exhibit 12, Panel B: Growth of $1 at an Equal One-Year Drawdown Budget

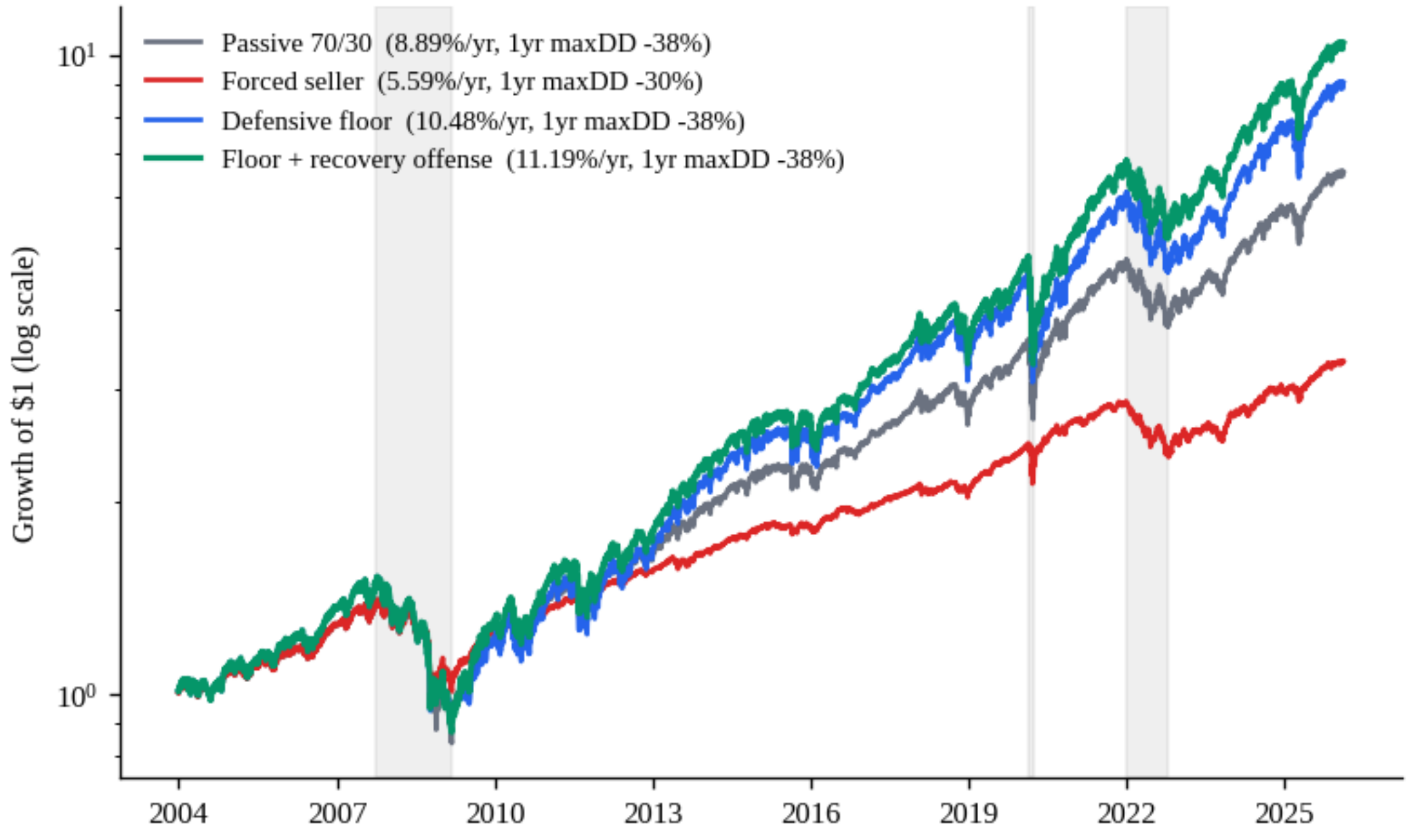

**Exhibit 14: Tracking-Error Signature of Offense, Defense, and the Combination**

| Policy | Compound return | One-year max drawdown | Tracking error | σ(TE) | Sharpe |
|---|---|---|---|---|---|
| Benchmark 70/30 | 8.89% | −38.2% | 0.00% | 0.00% | 0.543 |
| Offense (VIX lean, no floor) | 10.23% | −43.0% | 2.42% | 1.73% | 0.539 |
| Defensive floor | 10.48% | −38.2% | 5.30% | 4.33% | 0.540 |
| Floor + recovery offense | 11.19% | −38.2% | 5.12% | 3.38% | 0.556 |

## Online Appendix: Supplementary Exhibits

### Note on the Regime Signal

We use VIX percentiles rather than a Hamilton (1989) Markov-switching model for three reasons. The VIX is observable in real time and needs no estimation, whereas smoothed Markov state probabilities embed future information. Percentiles give a graduated signal, while switching probabilities cluster near zero or one and add turnover. And when we fit a two-state Markov-switching model to weekly S&P 500 returns and substitute it for the VIX regime, the two signals agree (Spearman 0.78, 84% concordance), but the Sharpe range across constraint levels widens from ±3% to ±14% as the binary signal concentrates risk-taking in fewer episodes.

**Exhibit 8, Panel A: Active Weight θ(t) Across Constraint Levels**

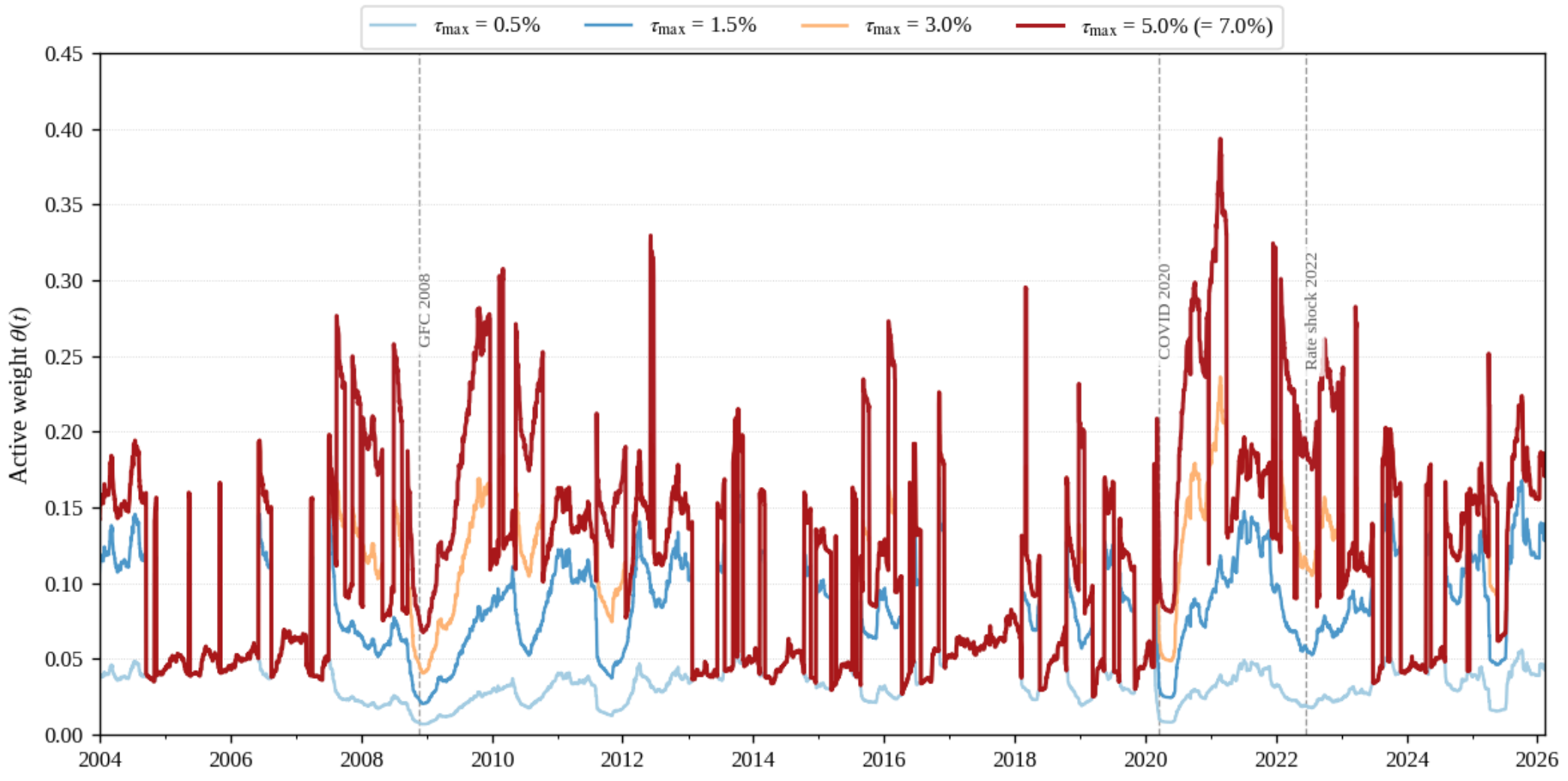

*0.5% binds 64.7% of days, 1.5% binds 64.7%, 3% binds 24.1%, 5% binds 0.0%. The 5% and 7% series coincide; only one is drawn.*

## Exhibit 8, Panel B: Realized Tracking Error Across Constraint Levels

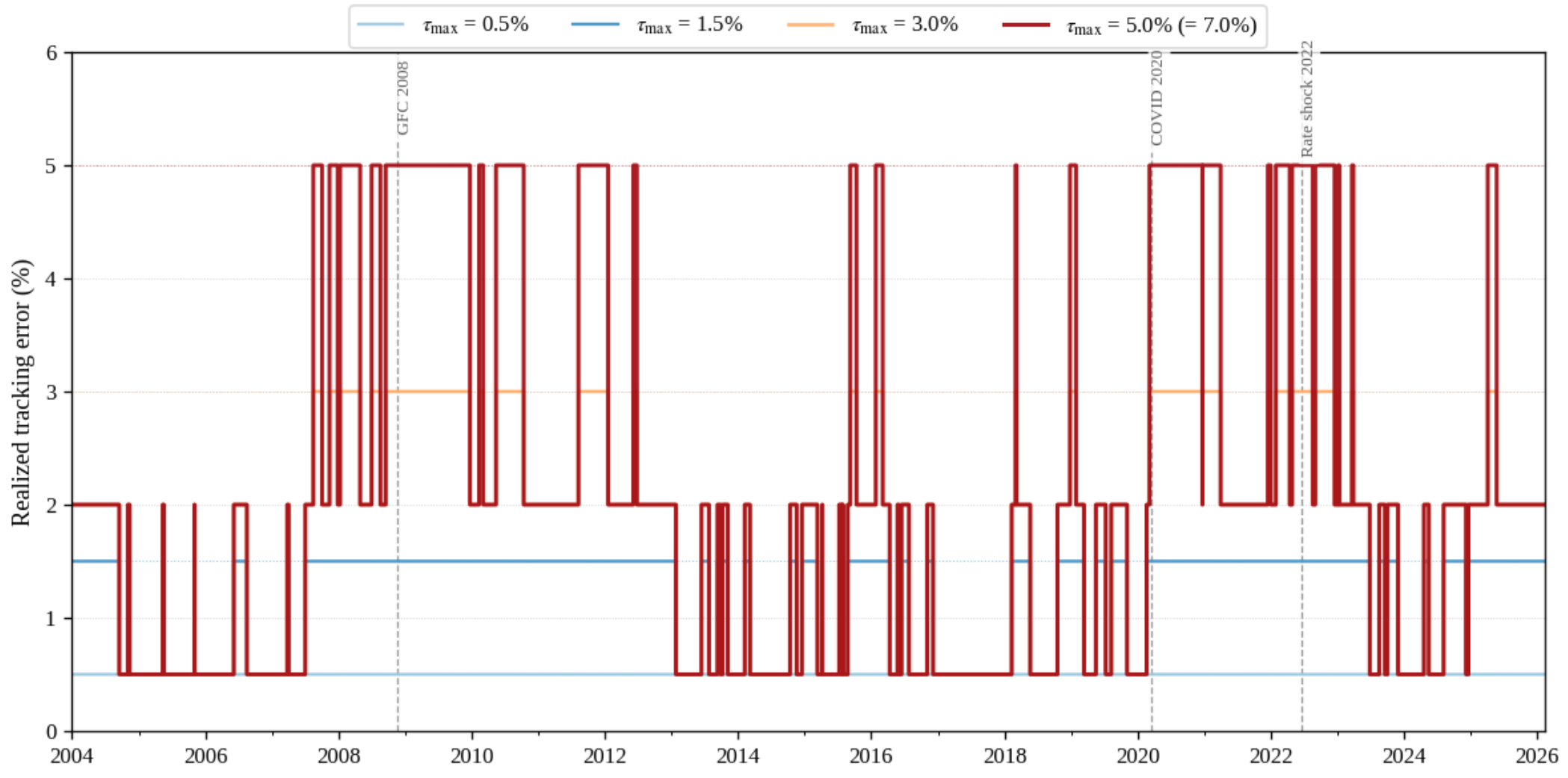

*Realized TE moves as a regime step function. Tight caps clip every mid- and high-VIX day; the 5% cap never binds.*

## Exhibit 13, Panel A: Base, Tracking-Error, and Dynamic Fund One-Year Drawdowns (2003–2026)

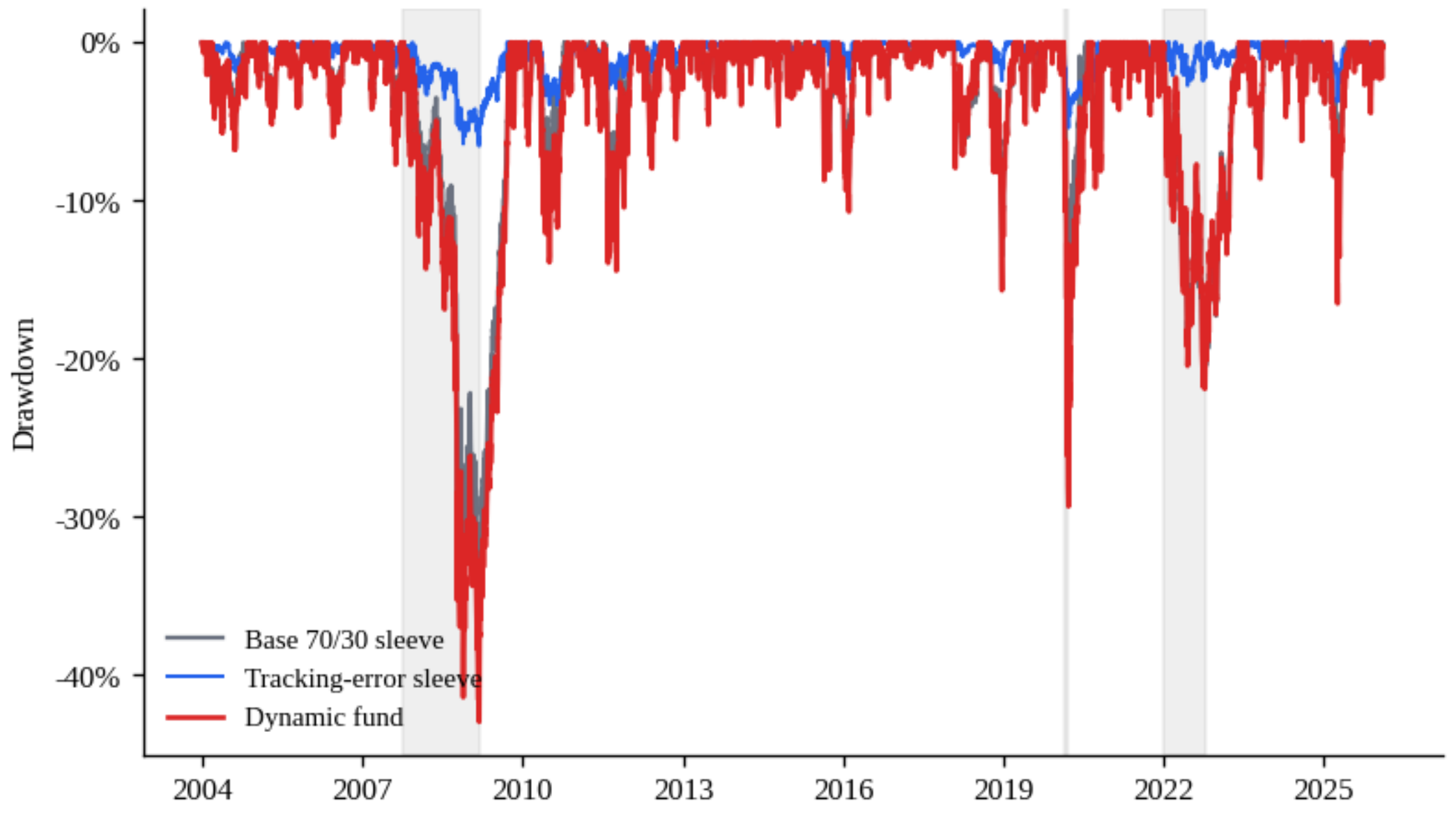

**Exhibit 13, Panel B: Base vs Tracking-Error Sleeve, Rolling 63-Day Return Correlation**

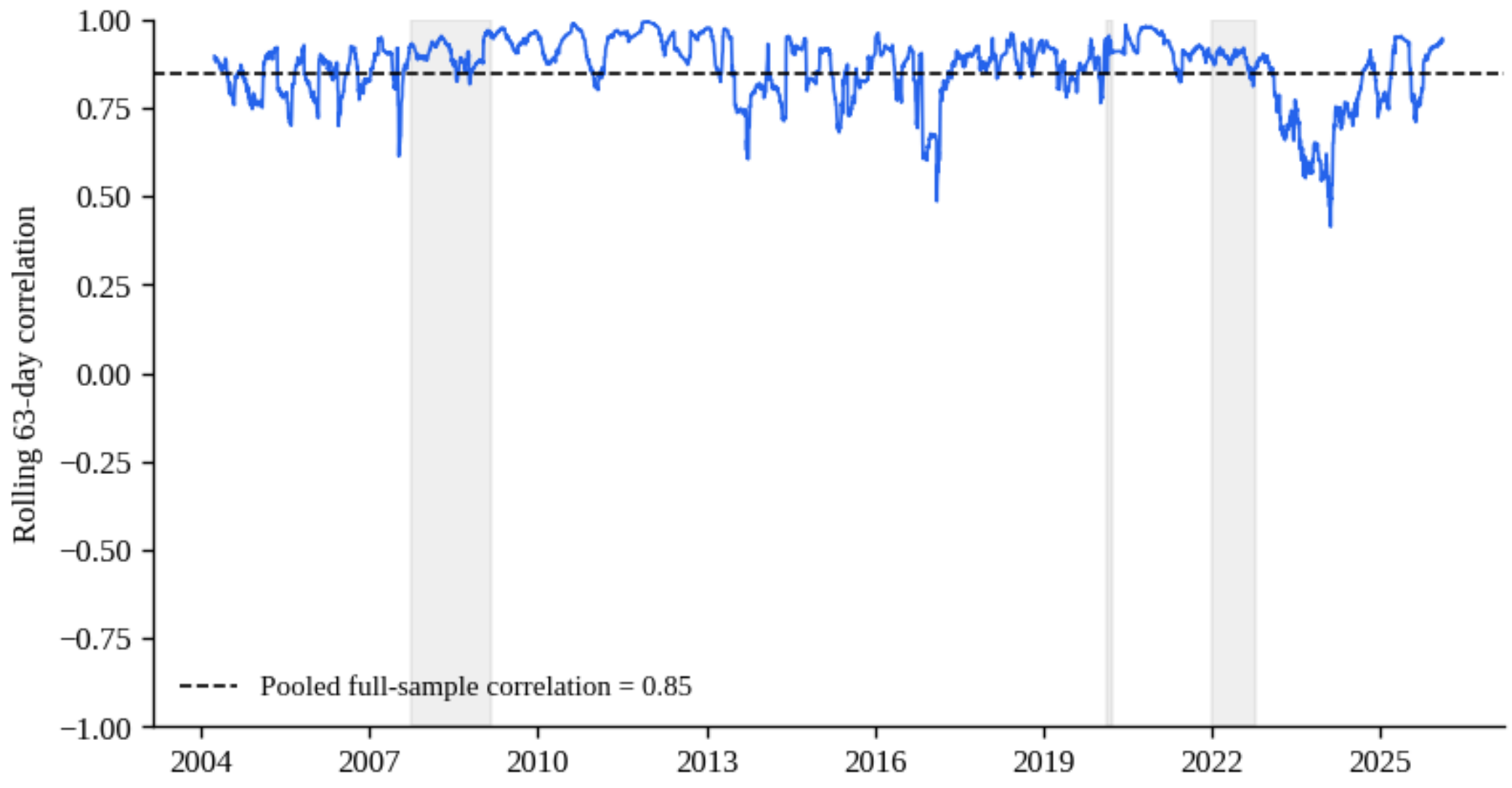